\documentclass[
preprint,
prd,
aps,
eqsecnum,
amsmath,
amssymb,
]{revtex4-1}
\def\slash#1{\, /\kern-0.6em{#1}}
\def\slashchar#1{\setbox0=\hbox{$#1$}           % set a box for #1 
   \dimen0=\wd0                                 % and get its size
   \setbox1=\hbox{/} \dimen1=\wd1               % get size of /
   \ifdim\dimen0>\dimen1                        % #1 is bigger
      \rlap{\hbox to \dimen0{\hfil/\hfil}}      % so center / in box
      #1                                        % and print #1
   \else                                        % / is bigger
      \rlap{\hbox to \dimen1{\hfil$#1$\hfil}}   % so center #1
      /                                         % and print /
   \fi}                                         %

\usepackage{graphicx}
\usepackage{color}
\usepackage{float}
\usepackage{subfloat}
\usepackage{subfigure}

\begin{document}

	\title{Quark ACM with topologically generated gluon mass} 
	\author{ Ishita Dutta Choudhury}\email{ishitadutta.choudhury@bose.res.in}
	\author{Amitabha Lahiri}\email{amitabha@bose.res.in}
	\affiliation{S. N. Bose National Centre For Basic Sciences,\\
		Block JD, Sector III, Salt Lake, Kolkata 700098, INDIA }

	\date{\today}
%	\maketitle
	
	\begin{abstract}
		We investigate the effect of a small, gauge-invariant mass of the gluon on the anomalous chromomagnetic moment of quarks (ACM) by perturbative calculations at one loop level. The mass of the gluon is taken to have been generated via a topological mass generation mechanism, in which the gluon acquires a mass through its interaction with an antisymmetric tensor field $B_{\mu \nu}$. For a small gluon mass $(<10$ MeV), we calculate the ACM at momentum transfer $q^2=-M_Z^2$. We compare those with the ACM calculated for the gluon mass arising from a Proca mass term. We find that the ACM of up, down, strange and charm quarks vary significantly with the gluon mass, while the ACM of top and bottom quarks show negligible gluon mass dependence. The mechanism of gluon mass generation is most important for the strange quarks ACM, but not so much for the other quarks. We also show the results at $q^2=-m_t^2$. We find that the dependence on gluon mass at $q^2=-m_t^2$ is much less than at $q^2=-M_Z^2$ for all quarks. 
\end{abstract}	
\maketitle		
		
%\keywords{Anomalous chromomagnetic moment; gluon mass; topological mass.}		
%	\end{abstract}
	
%\ccode{PACS No.s: }

%%%%%%%%%%%%%%%%%%%%%%%%%%
%%%%%%%%%%%%%%%%%%%%%%%%%%%%%%%
\section{Introduction}
%%%%%%%%%%%%%%%%%%%%%%%%%%%%%%%
The anomalous chromomagnetic moment (ACM) of quarks does 
not yet have a precise experimental bound. As the 
Large Hadron Collider (LHC) climbs new peaks of luminosity 
and energy, it opens new windows on precision QCD~\cite{Bernreuther:2015yna}, 
which should allow investigations into the anomalous couplings such as the 
ACM~\cite{Franzosi:2015osa}. 

The ACM of light quarks can be calculated in nonperturbative 
QCD~\cite{Kochelev:1996pv, Kochelev:2013zoa,  Kochelev:2013csa, Chang:2010hb}.
However, perturbative calculations of this quantity has received little 
attention so far, mainly because perturbative calculations in QCD do not 
make sense at very low energies. Since the 
QCD coupling constant diminishes rapidly with increasing energy, 
precision measurements are needed to measure the ACM at large 
momentum transfer. Although some bounds on top quark 
ACM have been provided by early theoretical analyses using a general 
effective Lagrangian with an anomalous coupling~\cite{ Rindani:2015vya,Martinez:1996cy, 
Martinez:2001qs, Martinez:2007qf, Gaitan:2015aia, Hioki:2009hm, Hioki:2010zu,
Kamenik:2011dk, Biswal:2012dr, Choudhury:2012np, 
Labun:2012ra, Degrande:2012gr, Ayazi:2013cba, Hioki:2013hva, 
Hesari:2014hva}, the only experimental analysis so far was done by the 
CMS collaboration last year~\cite{CMS:2014bea}. 

Another quantity not known very precisely, thus requiring
more investigation, is the mass of the gluon. Gluons are taken to be 
massless in QCD essentially because color symmetry is
unbroken as far as we know -- there is no spontaneous symmetry breaking
in QCD. However, thus far experiments have failed to find a 
stringent bound on the mass of the gluon~\cite{Beringer:1900zz}. 
Theoretical analyses regarding gluon mass have provided us with estimates 
varying over a large range, from a few MeV to several hundred 
MeV~\cite{Cornwall:1981zr, Aguilar:2009ke, 
	Parisi:1980jy, Field:2001iu, Yndurain:1995uq, Nussinov:2010jg}. 
However, if gluons are indeed massive particles, all gluons must have 
the same mass, since one gluon cannot be distinguished from another if
the SU(3) global symmetry is unbroken. There are different ways for 
a gluon to be massive without symmetry 
breaking~\cite{Cornwall:1981zr, Curci:1976bt, deBoer:1995dh}, the 
topological mass generation mechanism~\cite{Allen:1991gb, Hwang:1995er, 
	Lahiri:1996dm, Lahiri:2001uc} is one of them. 

In a previous 
paper~\cite{Choudhury:2014lna}, a small Proca mass was considered 
for the gluon, so that all gluons had the same mass, 
and the anomalous chromomagnetic moment of each quark
was calculated perturbatively. However, those results can be relevant 
only if the gluon is massive either via the Proca model, which is 
known to be unitary but non-renormalizable~\cite{Curci:1976bt, 
	deBoer:1995dh}, or gets a dynamically generated mass, in which 
case the mass is likely to go to zero at higher energies~\cite{Cornwall:1981zr} 
where perturbation calculations make any sense. The other possibility
is to use the topological mass mechanism, which does not break the 
symmetry, thus giving the same mass to all gluons, and may have the 
additional virtue of being unitary and renormalizable~\cite{Hwang:1995er, 
	Lahiri:1996dm, Lahiri:2001uc}. In this paper we present a 
perturbative calculation of the anomalous chromomagnetic moment 
of quarks, assuming a small, topologically generated, gauge-invariant 
mass for the gluon.

The quark-gluon interaction term which corresponds to the anomalous  
chromomagnetic moment is given by
\begin{equation}
ig\bar{\psi}  \sigma_{\mu \nu} T^{a} \psi\,  F_{2} (q^2) q^{\nu} G^{\mu a}\,.
\end{equation}
The coefficient $F_{2} (q^2)$ is the ACM of the quark at
momentum transfer $q$. We will calculate this term in perturbation theory at one loop, assuming a 
small mass for the gluon. Since QCD is a strongly coupled theory at low energies, the ACM cannot be 
calculated perturbatively for $q^2 = 0$, but only at sufficiently large momentum transfer. We will 
give numerical results for $q^2 = -M_Z^2$ as well as for $q^2 = -m_t^2$\,.

As mentioned above, a topological mass generation mechanism will be 
taken to be responsible for the mass of the gluon. This mechanism 
involves an antisymmetric tensor field $B_{\mu\nu}$ coupled to the 
field strength $F_{\mu\nu}$ of the gluon field through a $B\wedge F$ 
coupling. A Lagrangian which implements this is
\begin{equation}
{\cal L} =  -\frac{1}{4} F_{\mu \nu}^{a} F^{a\mu \nu} 
+ \frac{1}{12} H^{a}_{\mu \nu \lambda} H^{a\mu \nu \lambda}
+ \frac{M}{4} \epsilon ^{\mu \nu \rho \lambda}F^{a}_{\mu \nu}B^{a}_{\rho \lambda}\,,
\end{equation}
where $F_{\mu \nu}$  is the field strength tensor of the gluon field $G_\mu$, and
$H_{\mu\nu\lambda} = D_{[\mu}B_{\nu \lambda]}$ is the  field strength of the
antisymmetric tensor field $B^{a}_{\mu \nu}$. The gluon mass $M$ is a free 
parameter of the theory.

This Lagrangian is invariant under local $SU(3)$ gauge transformations
\begin{align}
	G_{\mu}  & \rightarrow U G_{\mu} U^{-1} -\frac{i}{g} \partial_{\mu} U U^{-1},\\
	B_{\mu \nu} & \rightarrow U B_{\mu \nu} U^{-1}\,.
\end{align}
For the quantization of the gauge fields we add two gauge fixing terms,
\begin{align}
	{\cal L}_{GF} = -\frac{1}{2\xi} (\partial_{\mu}G^{\mu}_{a})^2 - \frac{1}{2\eta} (D_{\mu}B^{\mu\nu}_{a})^2\,.
\end{align}
While the first of these two terms fixes the gauge for SU(3) transformations,
the role of the second term is somewhat more complicated. There is a higher
gauge transformation for the $B$ field, under $B_{\mu\nu} \to B_{\mu\nu} +
D_{[\mu}\Lambda_{\nu]}$\,,  which can be implemented in the Lagrangian by use
of an auxiliary field~\cite{Hwang:1995er, 	Lahiri:1996dm, Lahiri:2001uc}. 
This additional gauge transformation is fixed by the second 
term. We have not shown the auxiliary field because its couplings do not appear
in the one-loop diagrams responsible for the ACM. 

\begin{figure}[htbp]
	\centering
	\subfigure[]{\label{fig.vertex1_topo}
		\includegraphics[width=0.2\columnwidth]{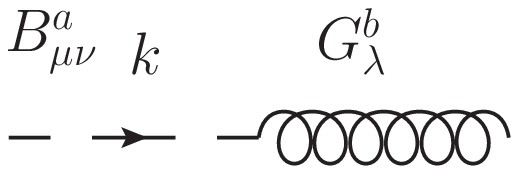}} 
	\hspace{1cm}          
	\subfigure[]{\label{fig.vertex2_topo}
		\includegraphics[width=0.2\columnwidth]{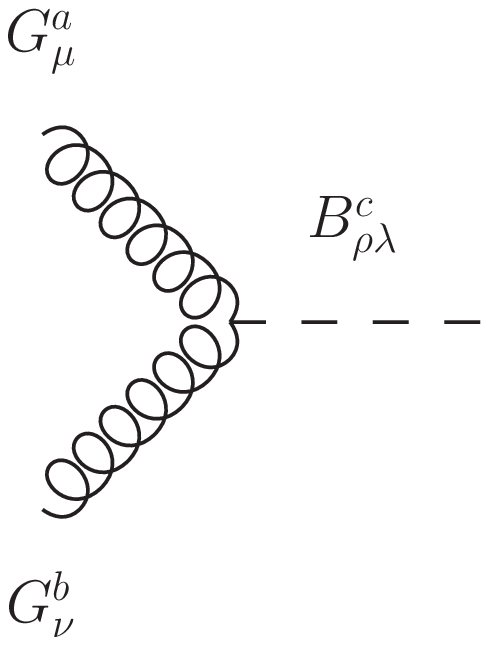}}
	\caption{(a) The  two-point $G-B$ vertex\,; (b) the cubic $G-G-B$ vertex}
	\label{GBGGB.fig}
\end{figure}

\begin{figure}[htbp]
	\centering
	\includegraphics[width=0.8\columnwidth]{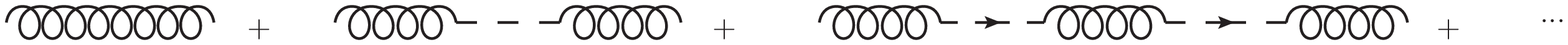}
	\caption{Bare gluon propagator by summing over all possible insertions of the $B$ propagator}
	\label{GBsum.fig}
\end{figure}

\begin{figure}[htbp]
	\centering
	\subfigure[]{\label{fig.vertex3_topo}
		\includegraphics[width=0.2\columnwidth]{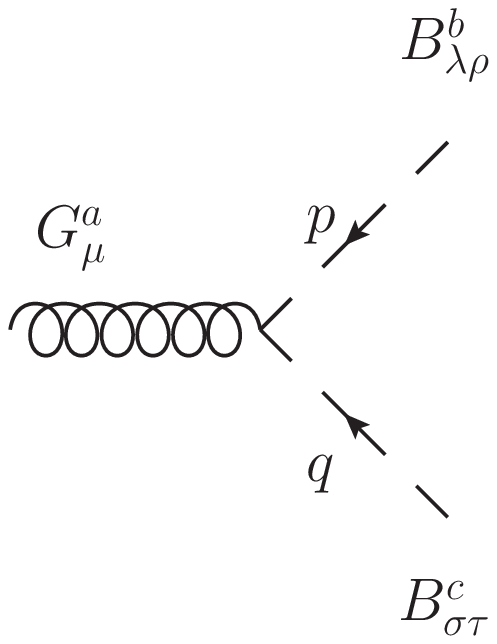}} 
	\hspace{1cm}               
	\subfigure[]{\label{fig.vertex4_topo}
		\includegraphics[width=0.2\columnwidth]{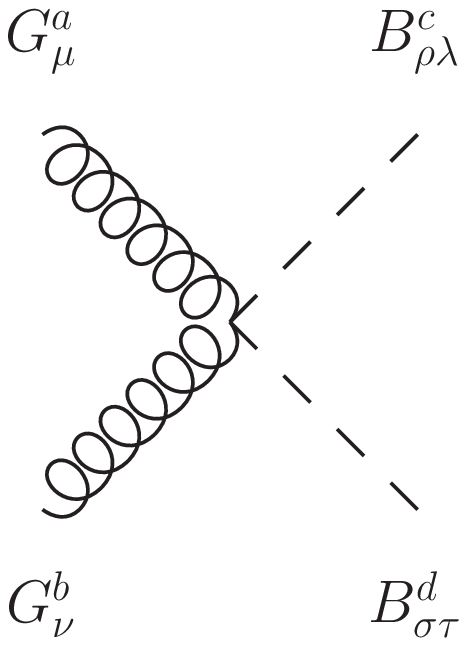}
	}
	\caption{(a) $GBB$ vertex and (b) $GGBB$ vertex, from $\dfrac{1}{12}H^a_{\mu\nu\lambda}H^{a\mu \nu \lambda}$} 
	\label{fig.vertex_topo}
\end{figure}

The propagators of the gluon and the tensor field $B_{\mu \nu}$ can now be calculated. 
Ignoring the mixed quadratic term for the moment, we find the propagators
\begin{align}
	i\Delta ^{\mu \nu , ab} &= -\frac{i}{k^2}\left(g^{\mu \nu} - (1-\xi)\dfrac{k^{\mu}k^{\nu}}
	{k^2}\right)\delta ^{ab},\\
	i\Delta ^{\mu \nu , \rho \lambda ; ab} &= \frac{i}{k^2}\left(g^{\mu[\rho}g^{\lambda]\nu} - (1-\eta)\dfrac{k^{\mu}k^{[\lambda}g^{\rho]\nu}-k^{\nu}k^{[\lambda}g^{\rho]\mu}}{k^2}\right)\delta ^{ab}\,.
\end{align}
The $B F$ interaction term contains a quadratic derivative interaction
between the two fields, as well as a cubic interaction. The vertices for 
these interactions are shown in Fig.~\ref{fig.vertex1_topo} and 
Fig.~\ref{fig.vertex2_topo}. The vertex rule corresponding to the 2-point
vertex diagram is 
\begin{align}
	iV^{ab}_{\mu \nu, \lambda} = - M \epsilon_{\mu \nu \lambda \rho} k^{\rho}\delta^{ab}\,.
	\label{V.GB}
\end{align}
The 2-point vertex corresponds to an off-diagonal mixing term between the 
gluon and the $B$ fields. The `effective' bare propagator of the gluon field 
is obtained by summing over the series of gluon propagators containing all 
possible insertions of the $B$ propagator via the 2-point vertex, as shown 
in Fig.~\ref{GBsum.fig}. The result is 
\begin{align}
	i D_{\mu \nu }= -i \dfrac{ g_{\mu \nu} - k_\mu k _\nu/k^2}{k^2-M^2 
		+i\epsilon} +i\xi \dfrac{k_\mu k_\nu	}{k^4}\,,
\end{align}
showing the presence of a pole in the propagator, corresponding to a mass 
$M$\,. We will see below that the terms proportional to $k_{\mu}k_{\nu}$ will 
not contribute in the ACM calculations, as all internal gluon lines couple 
to a conserved current at least at one end.
The three-point vertex of Fig.~\ref{fig.vertex2_topo} contributes to the one-loop 
diagrams for the ACM. The vertex rule for this diagram is 
\begin{equation}
i V^{ab}_{\mu, \nu ,\lambda \rho} = -igMf^{bca} \epsilon_{\mu \nu \lambda \rho}\,.
\label{V.GGB}
\end{equation}
There are two more vertices involving the $B$ field, coming from the $H^2$ term in the Lagrangian. These two are shown in Fig.~\ref{fig.vertex3_topo} and Fig.~\ref{fig.vertex4_topo}, and the corresponding vertex rules are
\begin{equation}
iV^{abc}_{\mu, \lambda\rho, \sigma\tau} =
gf^{abc}\left[(p - q)_\mu g_{\lambda[\sigma}g_{\tau]\rho} +
\left(p+q/{\eta}\right)_{[\sigma}g_{\tau][\lambda}g_{\rho]\mu}
-\left(q+p/{\eta} \right)_{[\lambda}g_{\rho][\sigma}g_{\tau]\mu}
\right]\,, \label{V.GBB}
\end{equation}
and
\begin{eqnarray}
\nonumber iV^{abcd}_{\mu, \nu, \lambda\rho, \sigma\tau} &=& ig^2
\left[f_{ace}f_{bde}\left(g_{\mu\nu}g_{\lambda[\sigma}g_{\tau]\rho}+ 
g_{\mu[\sigma}g_{\tau]g[\lambda}g_{\rho]\nu}-\frac{1}{\eta}
g_{\mu[\lambda}g_{\rho][\sigma}g_{\tau]\nu}\right) \right. 
\\  && \qquad+ \left.
f_{ade}f_{bce}\left(g_{\mu\nu}g_{\lambda[\sigma}g_{\tau]\rho} +
g_{\mu[\lambda}g_{\rho]g[\sigma}g_{\tau]\nu} -
\frac{1}{\eta}g_{\mu[\sigma}g_{\tau]g[\lambda}
g_{\rho]\nu}\right)\right].\qquad\,\label{V.GGBB}
\end{eqnarray}
Only the three-point vertex will contribute to the 1-loop diagrams for the ACM. 

There are of course other contributions to the ACM that we have to take into 
account. Let us list all the relevant 1-loop diagrams.
\begin{figure}[htbp]
	\centering
	\subfigure[]{\label{fig.qed}
		\includegraphics[width=0.18\columnwidth]{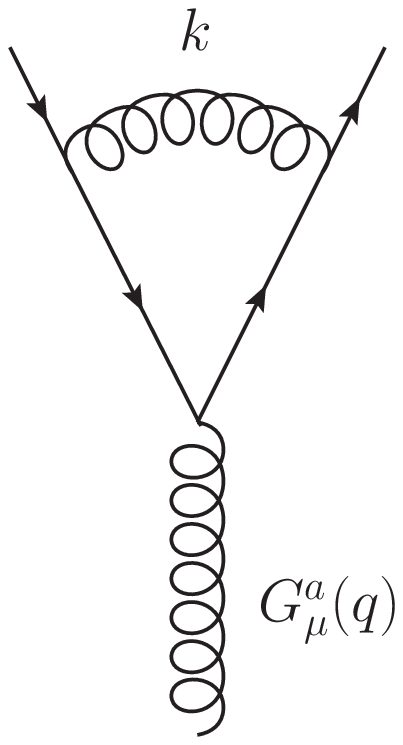}} \,
	%\hspace{1cm}               
	\subfigure[]{\label{fig.3g}
		\includegraphics[width=0.18\columnwidth]{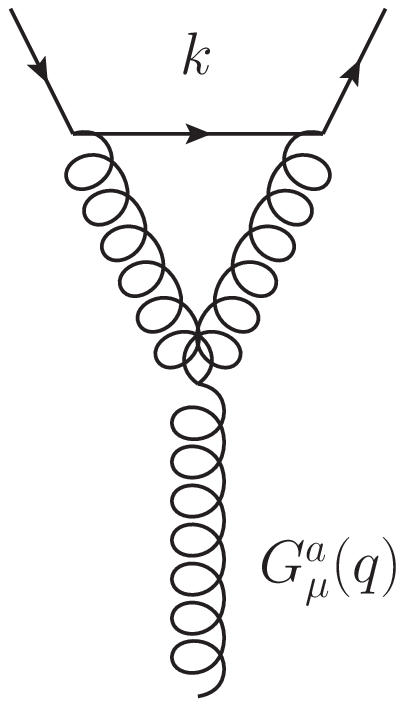}} \,
	%\hspace{1cm}
	\subfigure[]{\label{fig.eweak}
		\includegraphics[width=0.18\columnwidth]{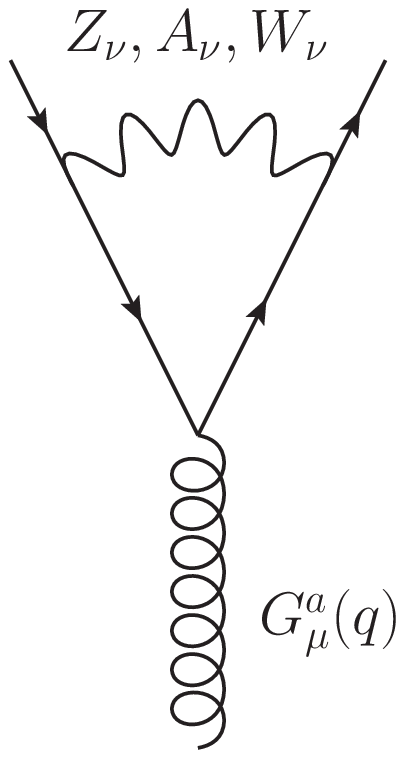}} \,
	%\hspace{1cm}               
	\subfigure[]{\label{fig.higgs}
		\includegraphics[width=0.18\columnwidth]{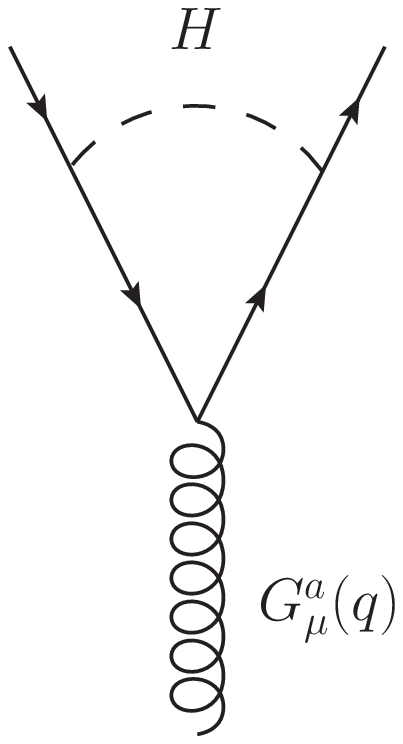}
	}
	\caption{Strong and electroweak contributions to the ACM of a quark:  strong contribution: (a) QED-like diagram; (b) purely 
		non-Abelian contribution; weak contributions: (c) gauge boson exchange;
		(d) Higgs boson exchange.} 
	\label{fig.strong_weak}
\end{figure}
The ACM of quarks receives contributions from both strong and weak processes at 
one loop order. Fig.~\ref{fig.strong_weak} shows the diagrams corresponding to strong 
and weak contributions. The new three-point vertices coming from topological mass 
generation give rise to some more diagrams which contribute to the gluon ACM. 
These new diagrams are shown in Fig.~\ref{fig.topo}.
\begin{figure}[htbp]
	\centering
	\subfigure[]{\label{fig.topo1}
		\includegraphics[width=0.2\columnwidth]{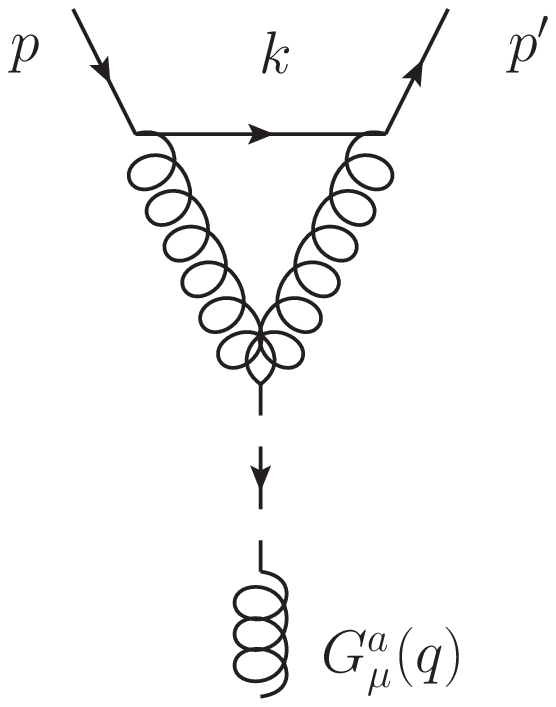}} 
	%\hspace{1cm}               
	\subfigure[]{\label{fig.topo2}
		\includegraphics[width=0.2\columnwidth]{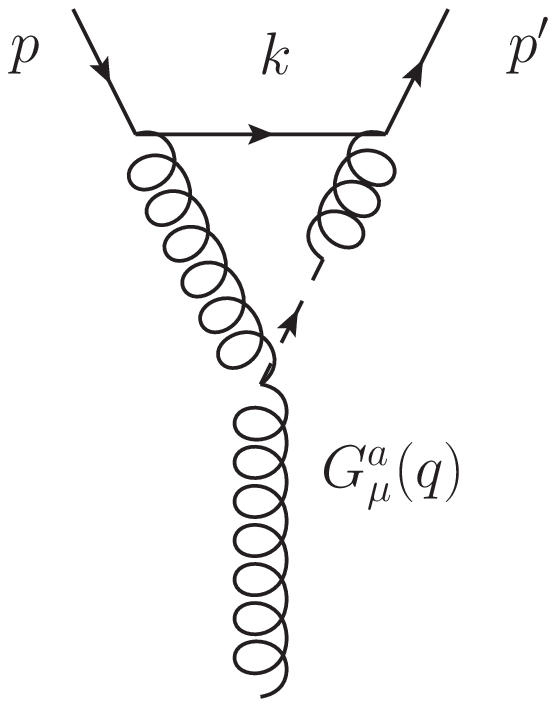}}
	\subfigure[]{\label{fig.topo3}
		\includegraphics[width=0.2\columnwidth]{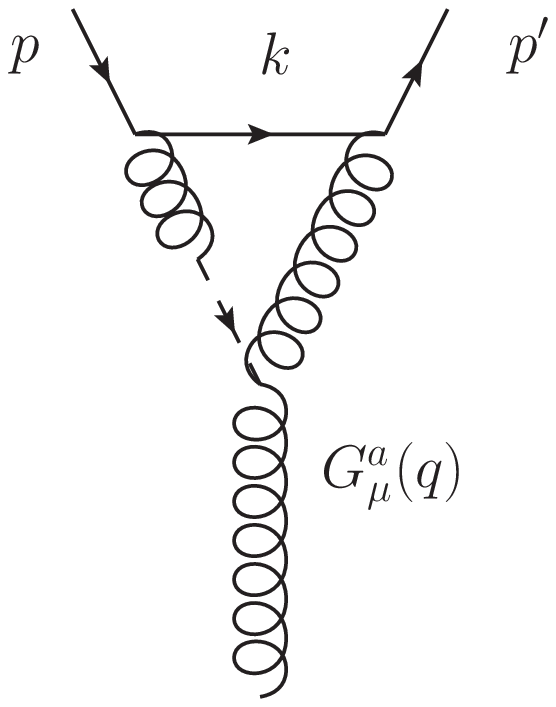}} 
	\\
	%\hspace{1cm}               
	\subfigure[]{\label{fig.topo4}
		\includegraphics[width=0.2\columnwidth]{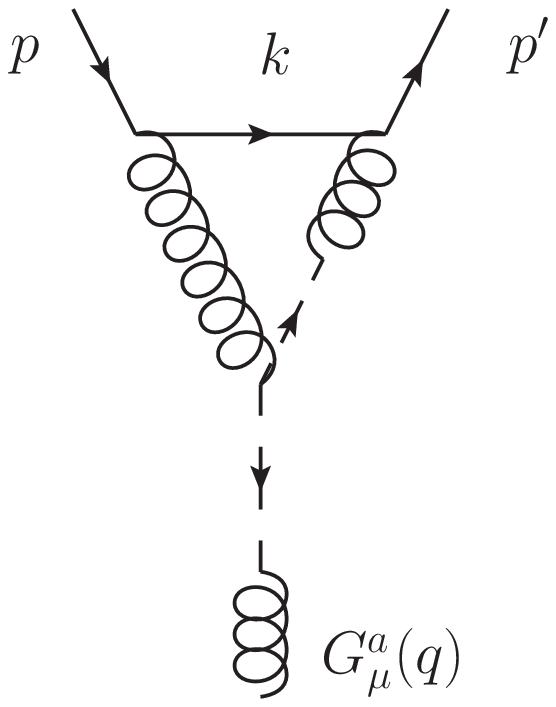}}
	\subfigure[]{\label{fig.topo5}
		\includegraphics[width=0.2\columnwidth]{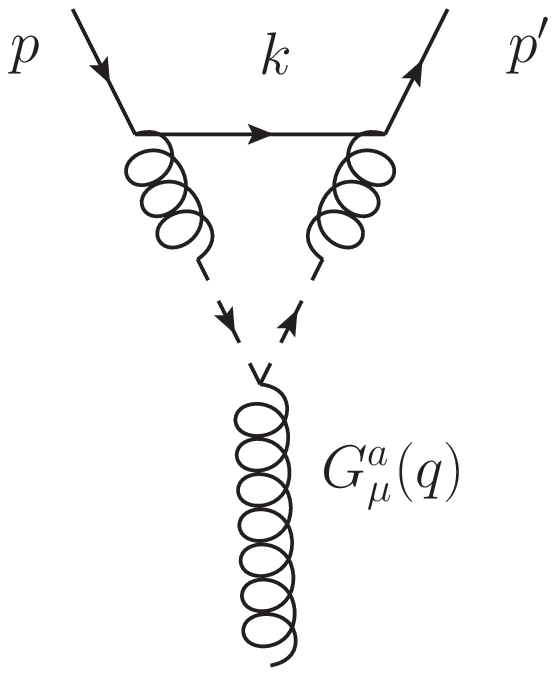}} 
	%\hspace{1cm}               
	\subfigure[]{\label{fig.topo6}
		\includegraphics[width=0.2\columnwidth]{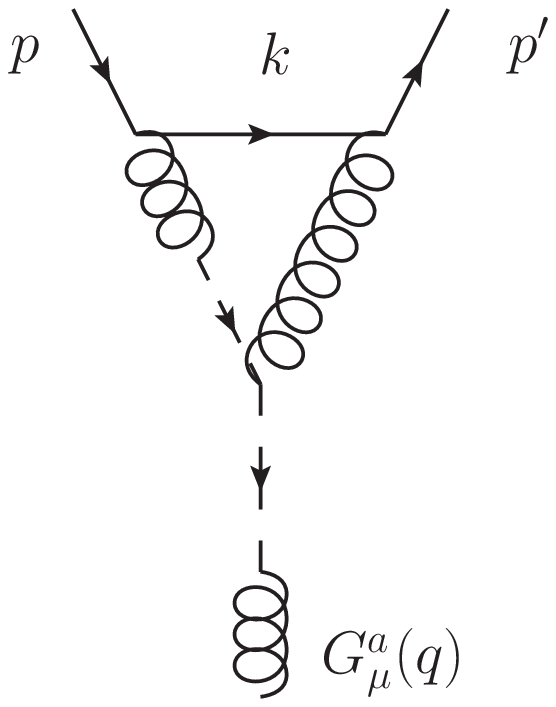}
	}               
	\caption{Topological contribution to the ACM of a quark.} 
	\label{fig.topo}
\end{figure}
Since $B$ does not couple to quarks directly, the contribution of the diagrams in 
Fig.~\ref{fig.topo} may be thought of as a correction to the diagram in Fig.~\ref{fig.3g}. 
We can calculate this conveniently by first removing the quark lines from each of the diagrams 
in Fig.~\ref{fig.topo} and then adding together the remaining parts. The resulting effective
topological three-gluon vertex can be written as 
%
%\warning{rewritten the LHS. does $V$ occur anywhere else?}
%
\begin{eqnarray}
V^{abc}(\bar{p},\bar{q},\bar{r}) &=&\frac{gM^2f^{abc}}{\bar{p}^{2}\bar{q}^{2}\bar{r}^{2}}\left[36\bar{q}^{2}\bar{r}^{2}\left(\bar{p}_{\mu}g_{\nu 
	\lambda}-\bar{p}_{\lambda}g_{\mu \nu}\right) 
%%%\right.\nonumber\\&& \left. 
+ 36\bar{p}^{2}\bar{q}^{2}\left(\bar{r}_{\nu}g_{\mu \lambda}-\bar{r}_{\mu}g_{\lambda \nu}\right) 
%%%\right.\nonumber\\&& \left. 
+ 36\bar{p}^{2}\bar{r}^{2}\left(\bar{q}_{\lambda}g_{\mu \nu}-\bar{q}_{\nu}g_{\lambda \mu}\right)\right.
\nonumber\\
&&\qquad+8\bar{q}^{2}\left[(\bar{r}-\bar{p})_{\mu} \bar{r}_{\nu}\bar{p}_{\lambda}-2\bar{p}_{\mu} \bar{r}_{\nu} \bar{r}_{\lambda}+2\bar{r}_{\mu}\bar{p}_{\nu}\bar{p}_{\lambda}   +(\bar{r} \cdot \bar{p})\lbrace (\bar{p}-\bar{r})_{\mu} g_{\lambda \nu}-2\bar{p}_{\nu} g_{\lambda \mu} + 2\bar{r}_{\lambda} g_{\mu \nu}
\rbrace\right]\nonumber\\
&& \qquad +8\bar{p}^{2}\left[(\bar{q}-\bar{r})_{\nu} \bar{r}_{\mu}\bar{q}_{\lambda}-2\bar{q}_{\mu} \bar{r}_{\nu} \bar{q}_{\lambda}+2\bar{r}_{\mu}\bar{q}_{\nu}\bar{r}_{\lambda}  +(\bar{q} \cdot \bar{r})\lbrace -(\bar{q}-\bar{r})_{\nu} g_{\lambda \mu}-2\bar{r}_{\lambda} g_{\mu \nu} + 2\bar{q}_{\mu} g_{\lambda \nu}
\rbrace\right]\nonumber\\
&& \qquad \left. +8\bar{r}^{2}\left[(\bar{p}-\bar{q})_{\lambda}\bar{p}_{\mu}\bar{q}_{\nu}-2\bar{p}_{\mu} \bar{p}_{\nu} \bar{q}_{\lambda}+2\bar{q}_{\mu}\bar{q}_{\nu}\bar{p}_{\lambda}   +(\bar{p} \cdot \bar{q})\lbrace (\bar{q}-\bar{p})_{\lambda} g_{\mu\nu}-2\bar{q}_{\mu} g_{\lambda \nu} - 2\bar{p}_{\nu} g_{\lambda \mu}
\rbrace\right]
\right]\,,\nonumber\\
\label{vertex.topo}
\end{eqnarray}
where all momenta are directed towards the vertex. This $V^{abc}$ is added to the usual three-gluon vertex, and the total is used to calculate Fig.~\ref{fig.3g}.

%%%%%%%%%%%%%%%%%%%%%%%%%%%%%%%%%%%%%%%%
\section{Calculations}
%%%%%%%%%%%%%%%%%%%%%%%%%%%%%%%%%%%%%%%%
We substitute $\bar{p} = p-k $ , $\bar{q} =-q $ and $\bar{r}= k-p^{\prime}$ in the total effective vertex given in Eq.~(\ref{vertex.topo}), and thus calculate the total contribution of all the diagrams in Fig.~\ref{fig.topo} to the vertex function. We can write the total contribution of the `topological' terms to the vertex function as
\begin{align}
	\Gamma_{\mu} 
	& = - 36 i M^2 \alpha_{s} \pi  \left[\Gamma^{(1)}_{\mu}+\Gamma^{(2)}_{\mu}+\Gamma^{(3)}_{\mu}
	+\frac{2}{9}\Gamma^{(4)}_{\mu}+\frac{2}{9} \Gamma^{(5)}_{\mu}
	+\frac{2}{9} \Gamma^{(6)}_{\mu}\right]\,.
	\label{Gamma}
\end{align}
The superscripts on the $\Gamma_{\mu}$ on the right hand side of Eq.~(\ref{Gamma}) 
correspond to the successive terms on the right hand side of Eq.~(\ref{vertex.topo}).
Using the relation 
\begin{equation}
T_{ji^{\prime}}^{c} T_{i^{\prime} i}^{b} f^{abc} = - \frac{i}{4} T_{ji}^{a}\,,
\end{equation}
we can write the different components of this equation as 
\begin{align}
	\Gamma^{(1)}_{\mu} &= \int \dfrac{d^4 k}{(2\pi)^{4}}\dfrac{(p-k)_{\mu} \gamma_{\lambda} (\slashchar{k} +m) \gamma ^{\lambda} -  (\slashchar{p}-\slashchar{k})(\slashchar{k} +m)\gamma_{\mu}}{(k^2-m^2)[(k-p)^2 -M^2][(k-p^{\prime})^2-M^2](k-p)^2 } \,,
	\label{Gamma1} \\
	%\end{align}
	%
	%\begin{align}
	\Gamma^{(2)}_{\mu} &= \int \dfrac{d^4 k}{(2\pi)^{4}}\dfrac{\gamma_{\mu} (\slashchar{k} +m)(\slashchar{k}- \slash{p^{\prime}}) - \gamma_{\lambda} (\slashchar{k}+m)\gamma^{\lambda} (k- p^{\prime})_{\mu}}{(k^2-m^2)[(k-p)^2 -M^2][(k-p^{\prime})^2-M^2](k-p^{\prime})^2 }\,,
	\label{Gamma2} \\
	%%\end{align}
	%%\begin{align}
	\Gamma^{(3)}_{\mu} &= \frac{1}{q^2} \int \dfrac{d^4 k}{(2\pi)^{4}}\dfrac{\gamma_{\mu}(\slashchar{k}+m)\slashchar{q}- \slashchar{q}(\slashchar{k}+m)\gamma_{\mu}}{(k^2-m^2)[(k-p)^2 -M^2][(k-p^{\prime})^2-M^2] q^2}\,,
	\label{Gamma3} 
\end{align}
\begin{align}
	\Gamma^{(4)}_{\mu}& =  \int \dfrac{d^4 k}{(2\pi)^{4}}\dfrac{1}{(k^2-m^2)[(k-p)^2 -M^2][(k-p^{\prime})^2-M^2](k-p)^2(k-p^{\prime})^2 }\nonumber\\
	& \qquad \times  \left [-(\slashchar{k}- \slashchar{p})(\slashchar{k}+ m)(\slashchar{k}-\slashchar{p^{\prime}})(2k-p-p^{\prime})_{\mu} +2(k-p)_{\mu} (\slashchar{k} - \slashchar{p^{\prime}})(\slashchar{k}+m)(\slashchar{k}-\slashchar{p^{\prime}})\right.\nonumber\\
	& \qquad + 2(k-p^{\prime})_{\mu} (\slashchar{k} - \slashchar{p})(\slashchar{k}+m)(\slashchar{k}-\slashchar{p})
	- (k-p^{\prime}) \cdot (k-p) \lbrace 2(\slashchar{k}- \slashchar{p^{\prime}})(\slashchar{k}+m) \gamma_{\mu}\nonumber\\
	& \qquad \left.- \gamma_{\lambda} (\slashchar{k}+m) \gamma^{\lambda} (2k-p-p^{\prime})_{\mu} + 2\gamma_\mu (\slashchar{k} +m) (\slashchar{k} - \slashchar{p})\rbrace\right] \,,
	\label{Gamma4}
\end{align}
\begin{align}
	\Gamma ^{(5)}_{\mu}& = \frac{1}{q^2} \int \dfrac{d^4 k}{(2\pi)^{4}}\dfrac{\gamma_{\lambda}(\slashchar{k}+m)\gamma_{\nu}g^{\nu\nu^{\prime}}g^{\lambda \lambda^{\prime}}}{ (k^2-m^2)[(k-p)^2 -M^2][(k-p^{\prime})^2-M^2](k-p^{\prime})^2 }\nonumber\\
	& \qquad \times   [q_{\lambda^{\prime}}(k-p^{\prime})_{\mu} (k+p-2p^{\prime})_{\nu^{\prime}} - 2(k-p^{\prime})_{\nu^{\prime}}q_{\lambda^{\prime}}q_{\mu} - 2q_{\nu^{\prime}}(k-p^{\prime})_{\lambda^{\prime}}(k-p^{\prime})_{\mu} \nonumber\\
	& \qquad- q \cdot (k-p^{\prime}) \lbrace g_{ \mu \lambda ^{\prime}}(p+k-2p^{\prime})_{\nu^{\prime}}-2g_{\mu \nu^{\prime}}(k-p^{\prime})_{\lambda^{\prime}}-2g_{\lambda^{\prime}\nu^{\prime}}
	 q_{\mu}\rbrace]\,,
	\label{Gamma5}
\end{align}
\begin{align}
	\Gamma ^{(6)}_{\mu}& = \frac{1}{q^2}  \int \dfrac{d^4 k}{(2\pi)^{4}}\dfrac{\gamma_{\lambda}(\slashchar{k}+m)\gamma_{\nu}g^{\nu\nu^{\prime}}g^{\lambda \lambda^{\prime}}}{ (k^2-m^2)[(k-p)^2 -M^2][(k-p^{\prime})^2-M^2](k-p)^2 }\nonumber\\
	& \qquad \times  [-(k-p)_{\mu} q_{\nu^{\prime}}(k-2p+p^{\prime})_{\lambda^{\prime}} - 2q_{\mu}(k-p)_{\lambda^{\prime}}q_{\nu^{\prime}} - 2q_{\lambda^{\prime}}(k-p)_{\nu^{\prime}}(k-p)_{\mu} \nonumber\\
	& \qquad +q \cdot (k-p) \lbrace g_{\mu \nu^{\prime}} (k+p^{\prime}-2p)_{\lambda ^{\prime}} - 2g_{\lambda^{\prime} \mu} (k-p)_{\nu^{\prime}} - 2g_{\nu^{\prime} \lambda^{\prime}}q_{\mu}\rbrace]\,.
	\label{Gamma6}
\end{align}
%
%%% where $\Gamma^a_{\mu}$, $\Gamma^b_{\mu}$,$\Gamma^c_{\mu}$,$\Gamma^d_{\mu}$,
%%% $\Gamma^e_{\mu}$ and $\Gamma^a_{\mu}$ correspond to the first, second, third, 
%%% fourth, fifth and sixth line respectively in Eq. (\ref{vertex.topo}).\

\begin{figure}[htbp]
	%\centering
	%\warning{the boxes should be of equal size!}
	\subfigure[]{\label{fig.topo_mz_u}
		\includegraphics[width=0.4\textwidth]{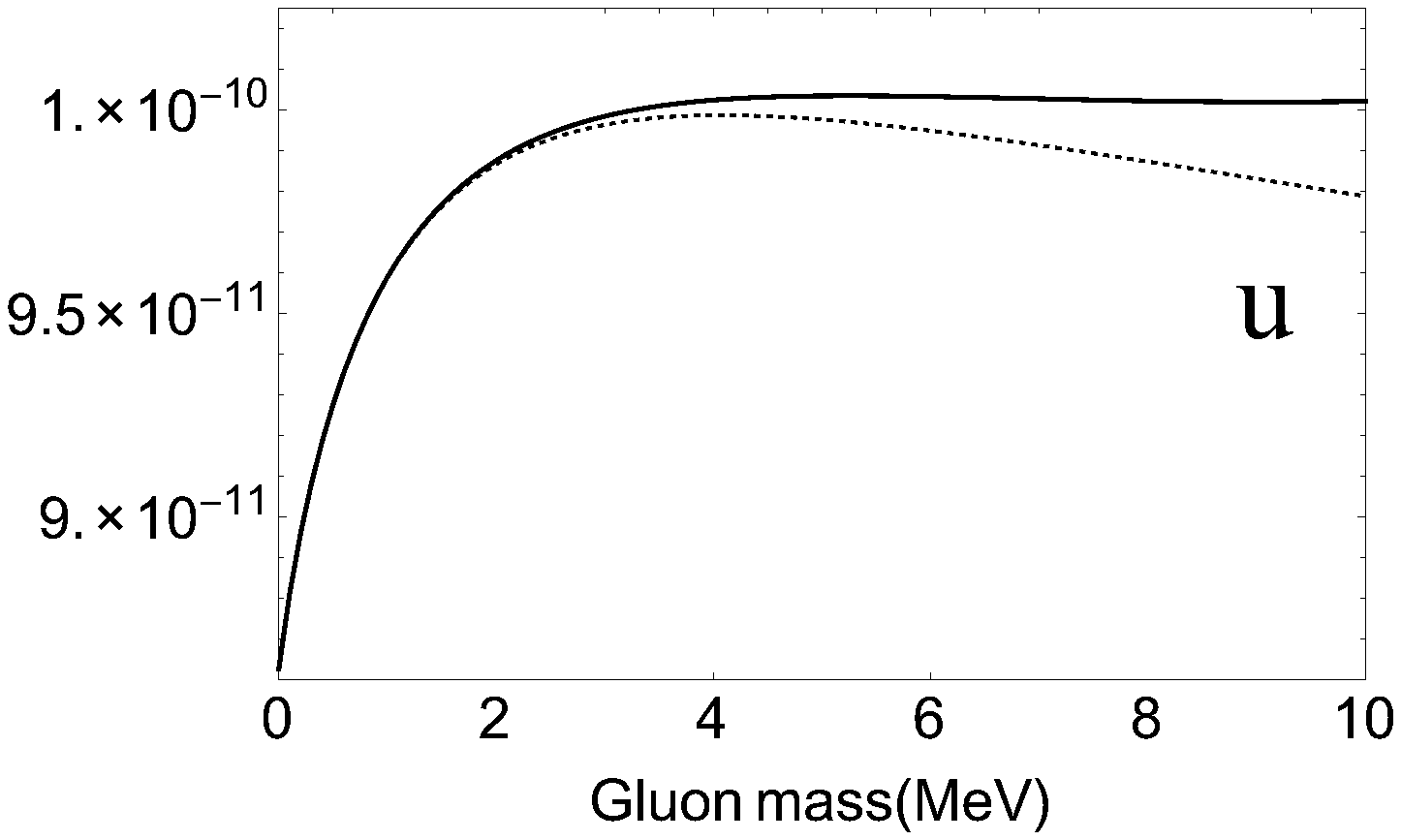}} 
	\hspace{0.1\textwidth}               
	\subfigure[]{\label{fig.topo_mz_d}
		\includegraphics[width=0.4\textwidth]{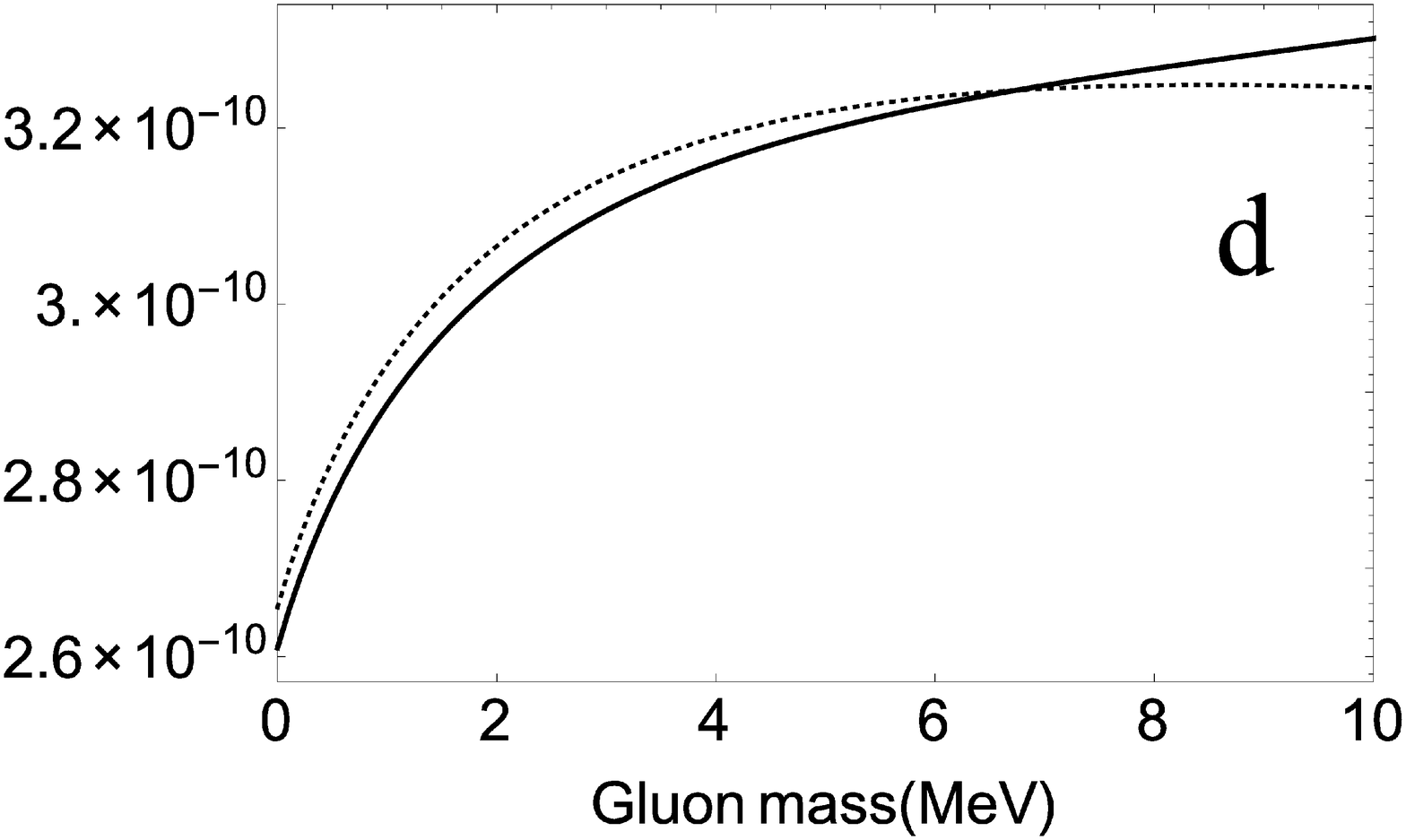}}\\
%	\hspace{0.02\textwidth}
	\subfigure[]{\label{fig.topo_mz_s}
		\includegraphics[width=0.4\textwidth]{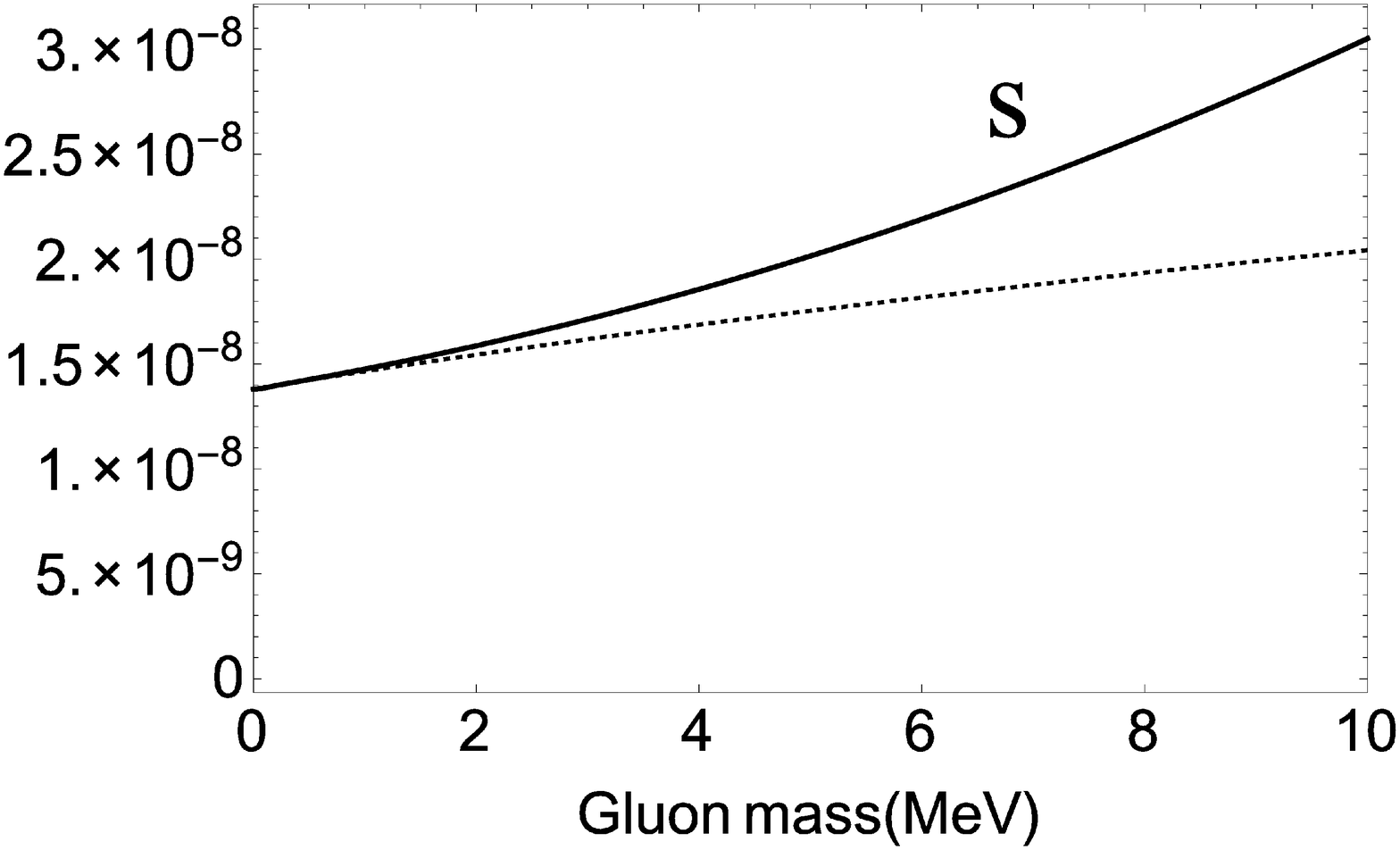}}
	\hspace{0.1\textwidth}
	\subfigure[]{\label{fig.topo_mz_c}
		\includegraphics[width=0.4\textwidth]{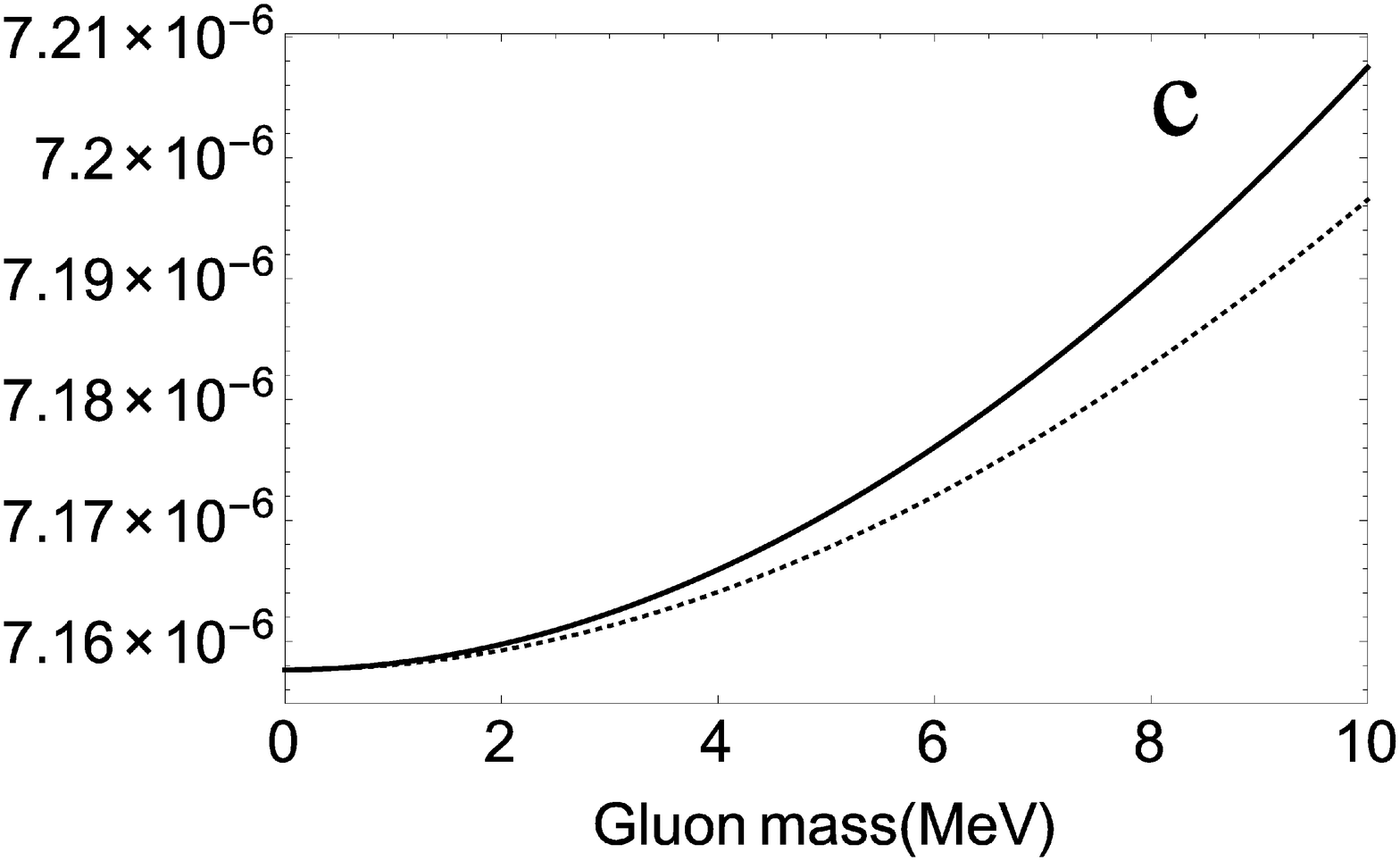}}\\
%	\hspace{0.02\textwidth}
	\subfigure[]{\label{fig.topo_mz_b}
		\includegraphics[width=0.4\textwidth]{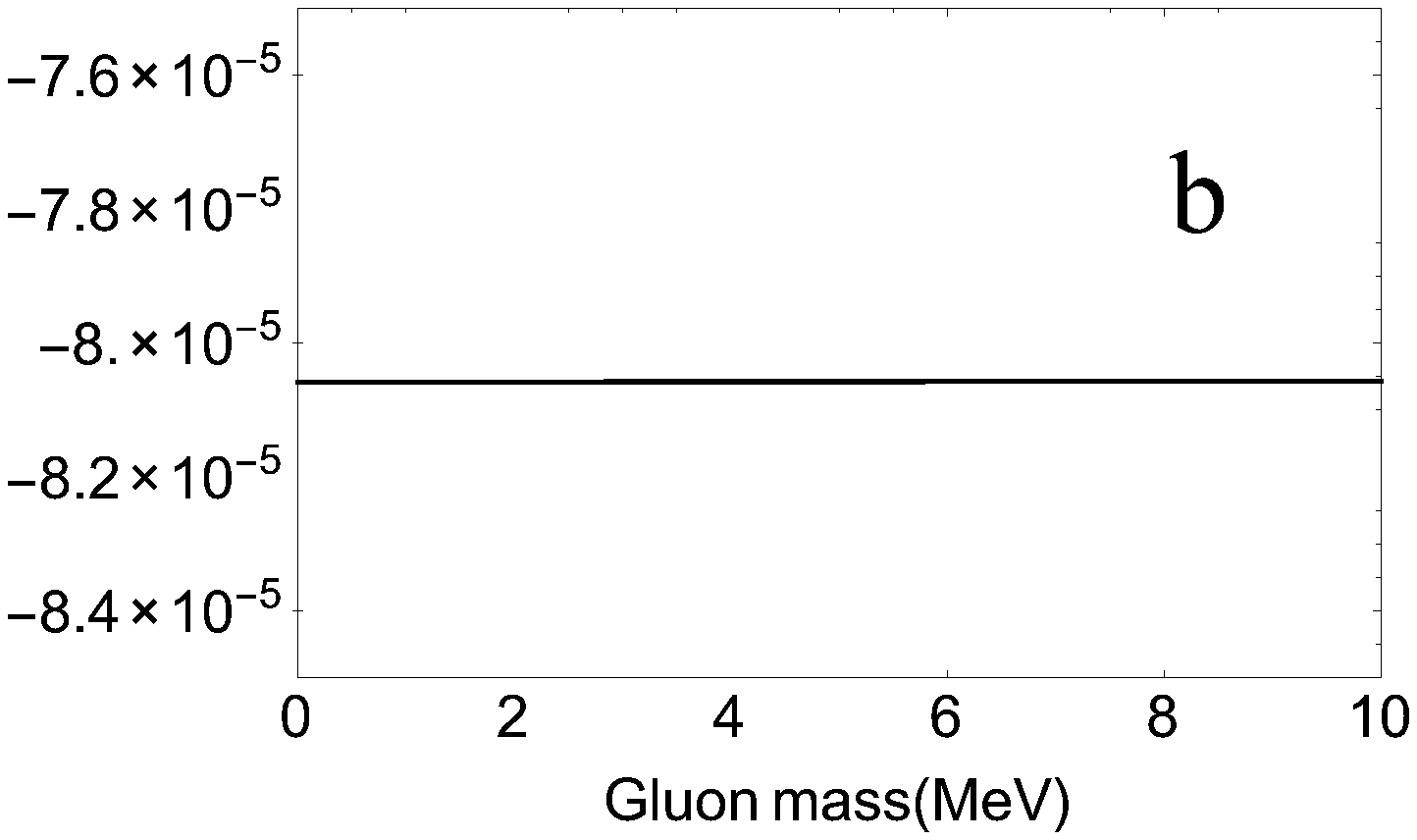}}
	\hspace{0.1\textwidth}
	\subfigure[]{\label{fig.topo_mz_t}
		\includegraphics[width=0.4\textwidth]{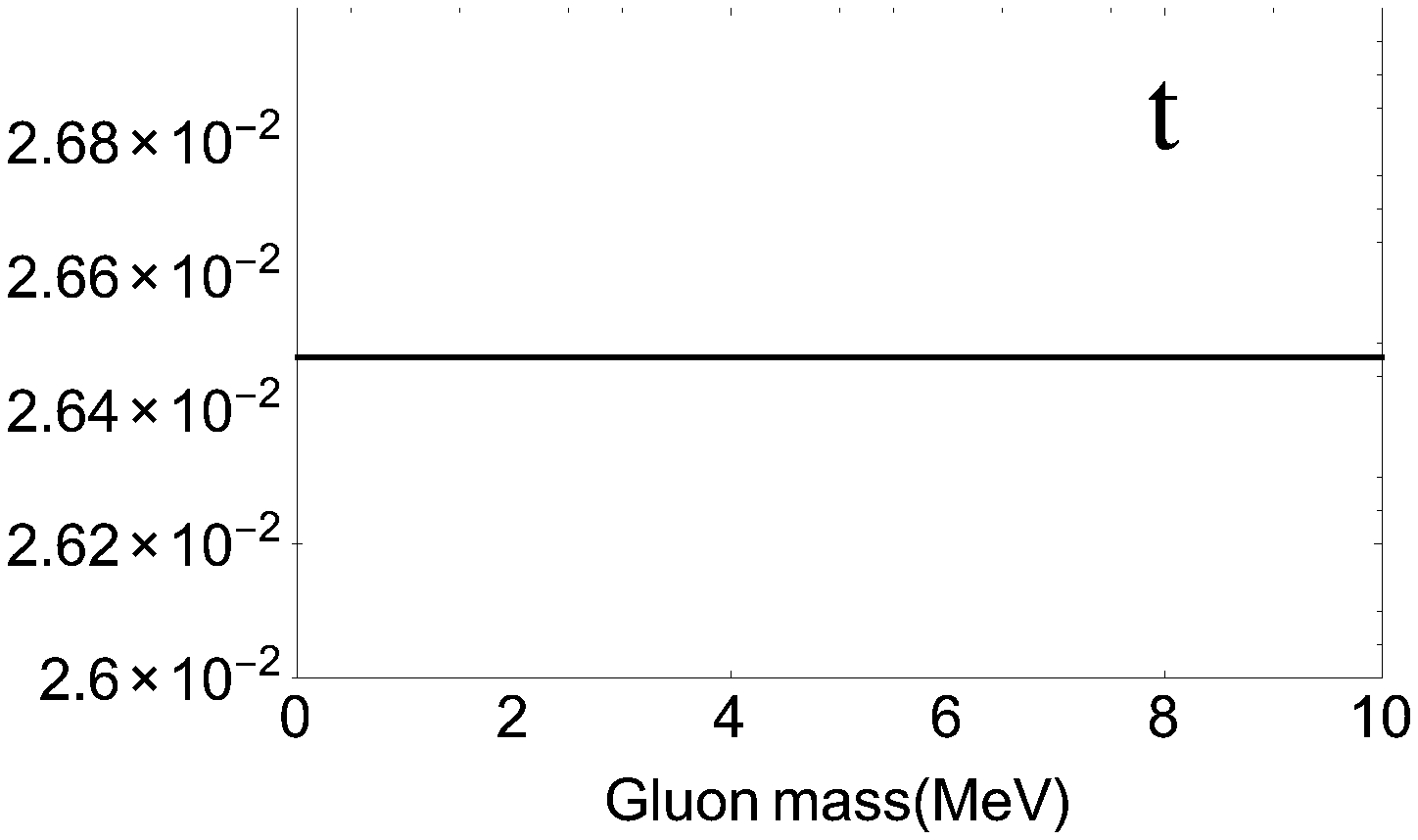}}
	\caption{$4mF_2$ of quarks at $q^2 = -M^2_Z$\,; continuous lines represent dependence 
		on topologically generated gluon mass; dotted lines represent dependence on gluon mass 
		coming from a Proca term.} 
	\label{fig.total_topo_mz}
\end{figure}

\begin{figure}[htbp]
%	\centering
	%\warning{the boxes should be of equal size!}
	\subfigure[]{\label{fig.topo_mt_u}
		\includegraphics[width=0.4\textwidth]{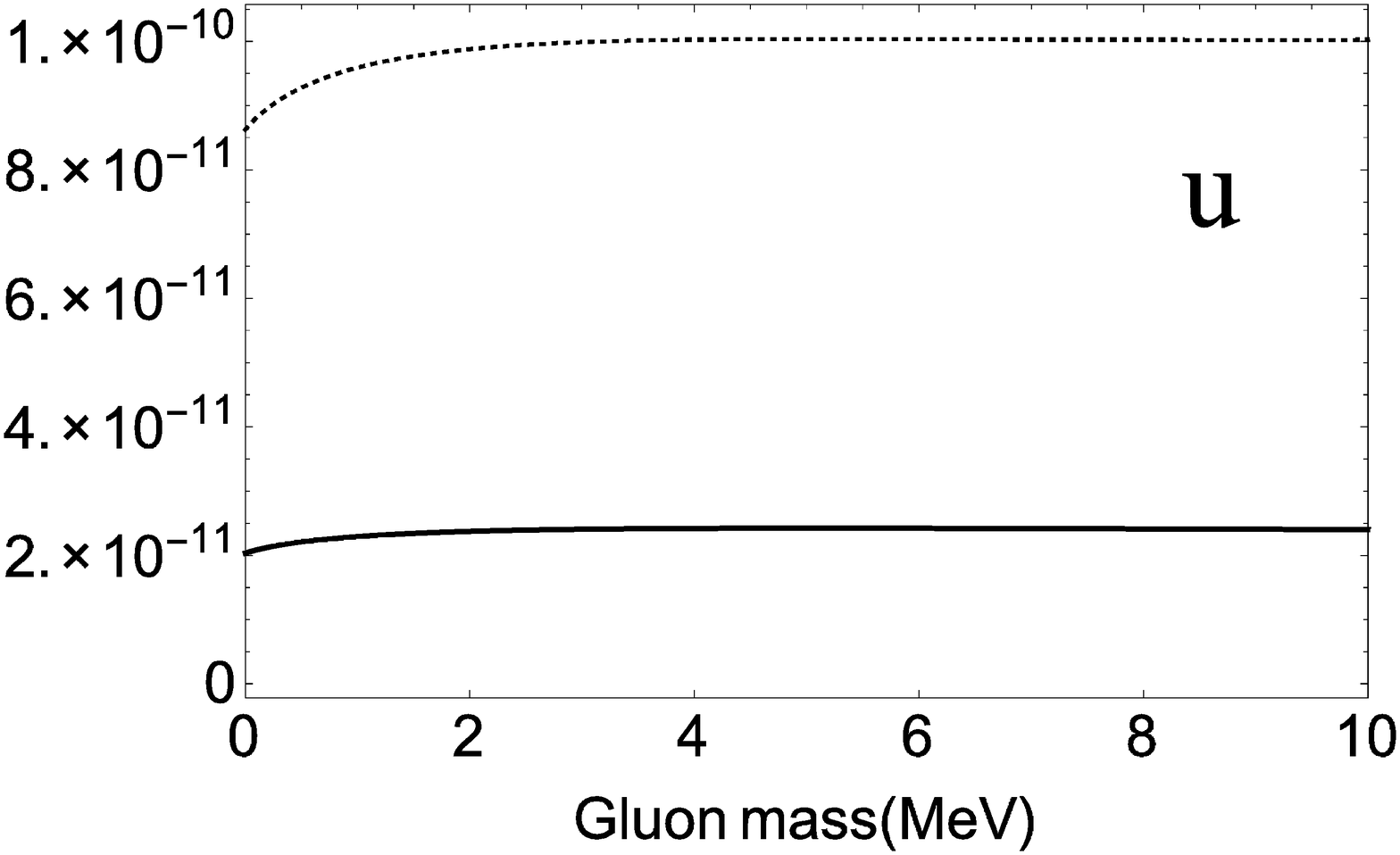}} 
	\hspace{0.1\textwidth}               
	\subfigure[]{\label{fig.topo_mt_d}
		\includegraphics[width=0.4\textwidth]{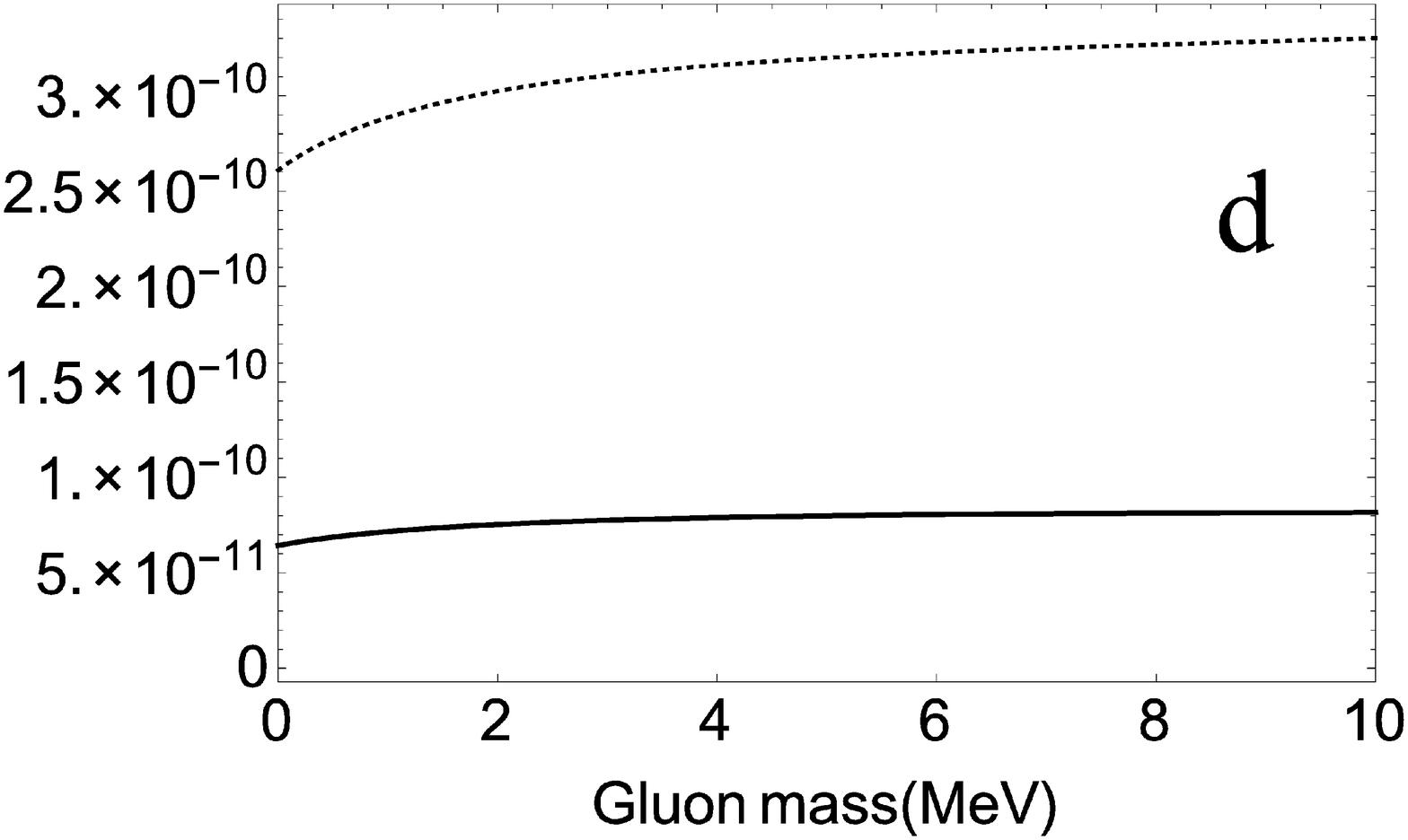}}\\
	%\hspace{0.02\textwidth}
	\subfigure[]{\label{fig.topo_mt_s}
		\includegraphics[width=0.4\textwidth]{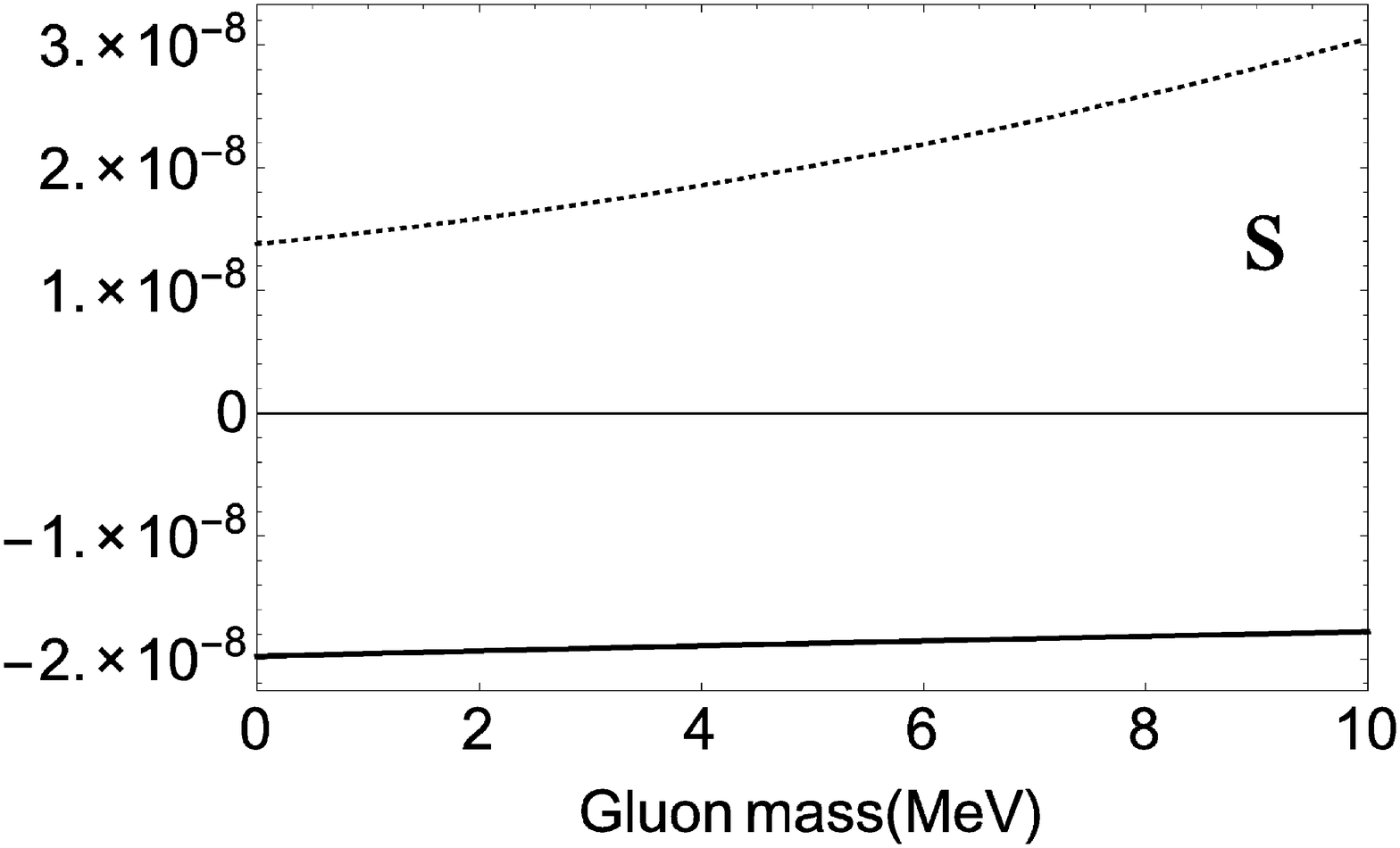}}
	\hspace{0.1\textwidth}
	\subfigure[]{\label{fig.topo_mt_c}
		\includegraphics[width=0.4\textwidth]{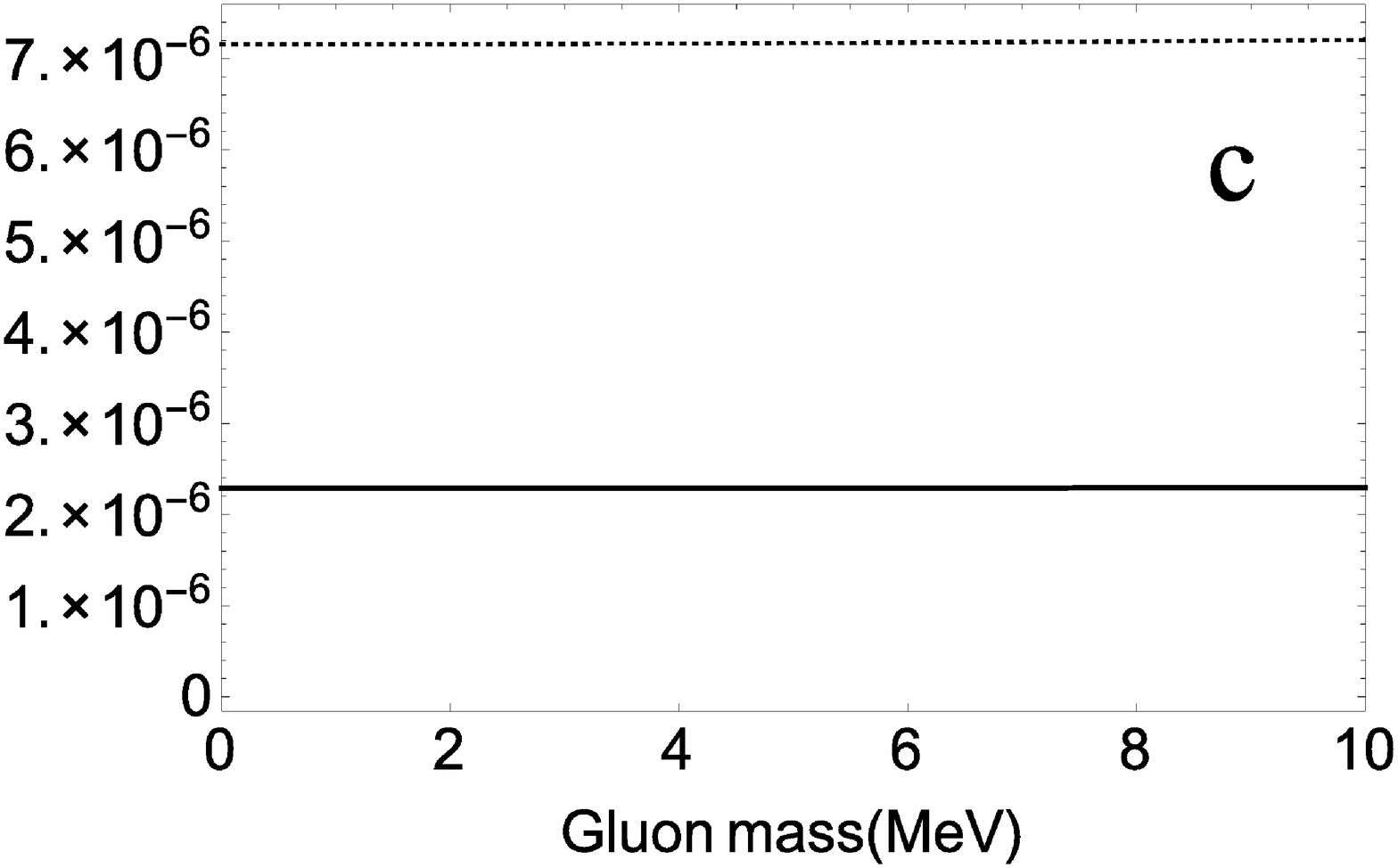}}\\
%	\hspace{0.02\textwidth}
	\subfigure[]{\label{fig.topo_mt_b}
		\includegraphics[width=0.41\textwidth]{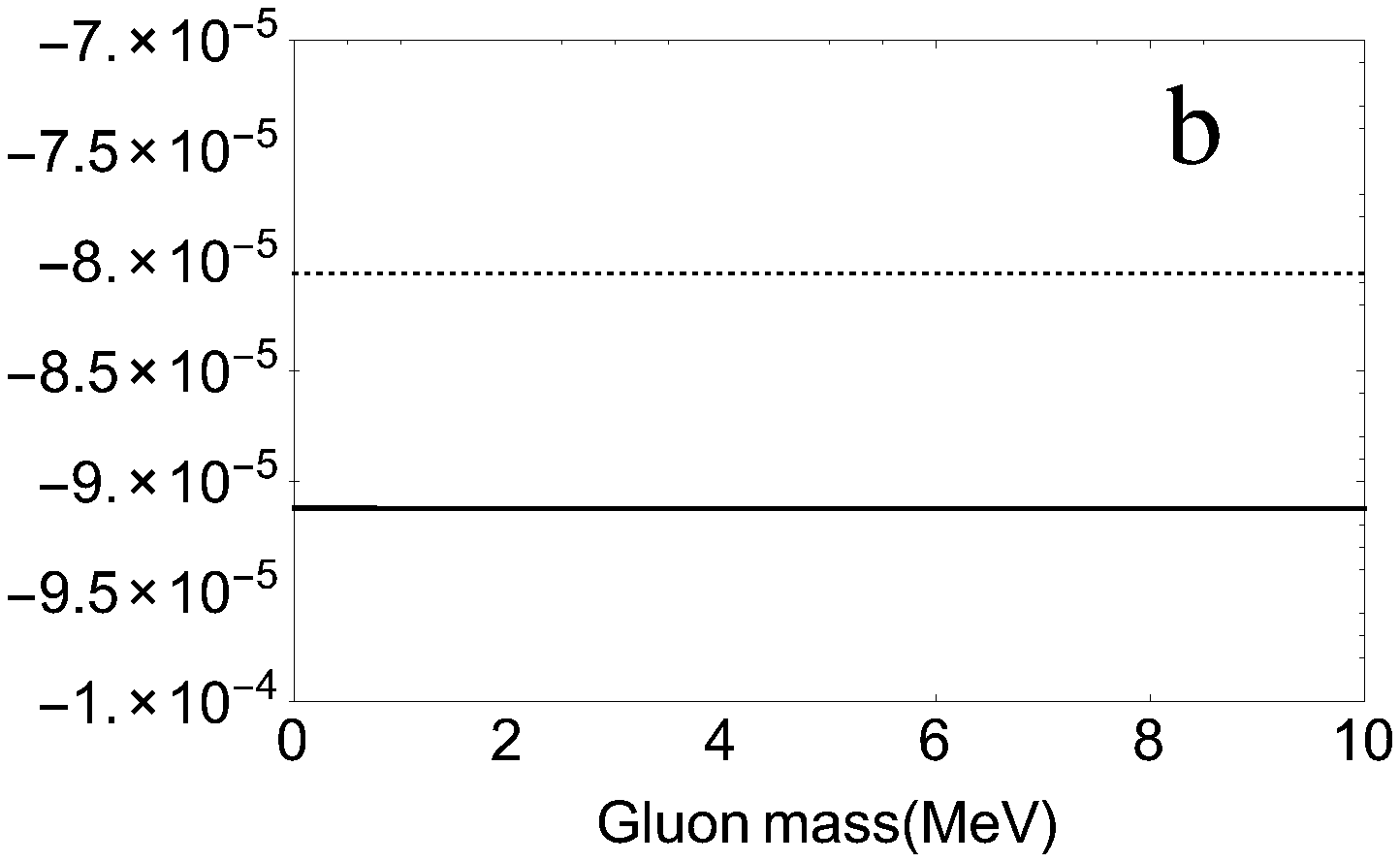}}
	\hspace{0.1\textwidth}
	\subfigure[]{\label{fig.topo_mt_t}
		\includegraphics[width=0.41\textwidth]{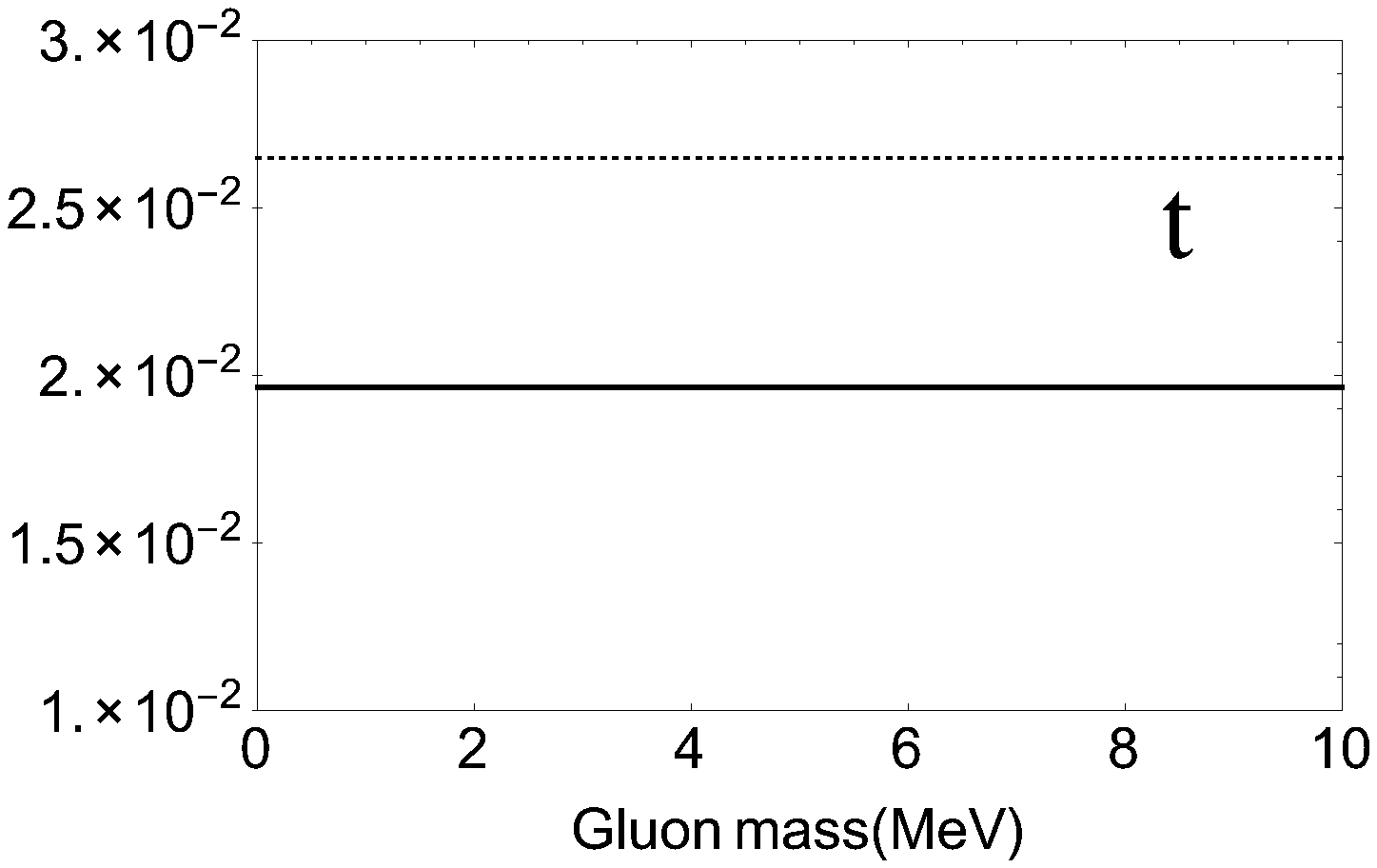}}
	\caption{$4mF_2$ of quarks ; continuous lines represent dependence 
		on topologically generated gluon mass at $q^2 = -m^2_t$\,; dotted lines represent dependence on gluon mass at $q^2 = -M^2_z$\,.} 
	\label{fig.total_topo_mz_mt}
\end{figure}

The calculation of the contributions from these terms to the ACM are given 
in the Appendix. Adding Eqs.~(\ref{F2q^2.topoa+b}),~(\ref{F2Q^2.topoc}),~(\ref{F2q^2.topod})
and~(\ref{F2q^2.topoe+f}), we get
\begin{align}
	F^{TM}_{2}({q^2 = -M_Z^2})
%	{(q^2 = -M_Z^2)}
	& = \dfrac{M^2  \alpha_{s}}{m^2\pi} \left[ 2 \int \limits_{0}^{1} d\zeta_{1}\!\int \limits_{0}^{1-\zeta_{1}} 
	d\zeta_{2} \int \limits_{0}^{1-\zeta_{1}-\zeta_{2}} d\zeta_{3}\, \dfrac{\zeta_{1} (\frac{9}{4}\zeta_{1} 
		+ \zeta_{3})}{\left[\zeta_{1}^2 +\zeta_{3}(1-\zeta_{1}-\zeta_{3})\frac{M_{Z}^{2}}{m^2}+(\zeta_{2}
		+\zeta_{3})\frac{M^{2}}{m^2}\right]^2}\right. \nonumber\\
	& \qquad - \int \limits_{0}^{1} d\zeta_{1}\int \limits_{0}^{1-\zeta_{1}} d\zeta_{2} \int \limits_{0}^{1-\zeta_{1}-\zeta_{2}} d\zeta_{3}  \int \limits_{0}^{1-\zeta_{1}-\zeta_{2}-\zeta_{3}} d\zeta_{4}\nonumber\\
	& \qquad \times \dfrac{\zeta_{1}(1-2\zeta_{1})\lbrace \zeta_{1}^2 +(\zeta_{2}+\zeta_{3})(1-\zeta_{1}-\zeta_{2}-\zeta_{3})\frac{M_{Z}^2}{m^2}-\zeta_{1}^3\frac{M_{Z}^2}{m^2} +3\zeta_{1}(\zeta_{2}+\zeta_{4})\rbrace}{\left[\zeta_{1}^2+ (\zeta_{2}+\zeta_{3})(1-\zeta_{1}-\zeta_{2}-\zeta_{3})\frac{M_{Z}^2}{m^2} +(\zeta_{2}+\zeta_{4})\frac{M^2}{m^2}\right]^3}\nonumber\\
	&\qquad \left. - \frac{9M_{Z}^2}{2m^2} \int \limits_{0}^{1} d\zeta_{1}\int \limits_{0}^{1-\zeta_{1}} d\zeta_{2} \dfrac{\zeta_{1}}{\zeta_{1}^2+\zeta_{2}(1-\zeta_{1}-\zeta_{2})\frac{M_{Z}^2}{m^2} + (1-\zeta_{1})\frac{M^2}{m^2}}\right]\,.
\end{align}
This is the contribution due to the topological mass mechanism, i.e., due to the diagrams in
Fig.~\ref{fig.topo}. To this we need to add the contributions due to the strong and the weak
interactions. The calculations for both of these are exactly as given in~\cite{Choudhury:2014lna}. 
The contribution of the diagram in Fig.~\ref{fig.qed} to a quark of mass $m$ is
\begin{align}
	F^{\tiny(\ref{fig.qed})}_{2}(q^2 = -M_Z^{2}) = -\dfrac{\alpha_{s}}{12\pi m}\int\limits_{0}^{1}d\zeta_{3}\int
	\limits_{0}^{1-\zeta_{3}}d\zeta_{2}\dfrac{(1- \zeta_{3})
		\zeta_{3}}
	{ (1- \zeta_{3})^{2}  
		+ (1-\zeta_{2}-\zeta_{3})\zeta_{2}\left(\frac{M_{Z}}{m}\right)^2 + 
		\zeta_{3}\left(\frac{M}{m}\right)^2}\,,
	\label{F2(q^2).qed}
\end{align}
and that of the diagram in Fig.~\ref{fig.3g} is
\begin{align}
	F_{2}^{\tiny(\ref{fig.3g})} (q^2 = -M_Z^{2})  =\dfrac{\alpha_{s}}{8\pi m}\int\limits_{0}^{1}d\zeta_{3}\int
	\limits_{0}^{1-\zeta_{3}}d\zeta_{2}\,\dfrac{(1- \zeta_{3})
		\zeta_{3}}
	{ \zeta_{3}^{2}  
		+ (1- \zeta_{2}-\zeta_{3})\zeta_{2}(\frac{M_Z}{m})^{2} 
		+(1- \zeta_{3})\left(\frac{M}{m}\right)^{2}}\,.
	\label{F2(q_^2).3g}
\end{align}
\begin{table} [htbp]
	\centering
	%\begin{tabular}{c p{12 mm} c p{15 mm} c p{16 mm} c p{12 mm} c p{12
	%mm} c p{12 mm}} 
	\caption{Electroweak contribution to $4m F_{2}(q^{2}= - M_Z^{2})$ 
		of each quark}{
	\begin{tabular}{cccccc}
		\hline\hline
		Quark & $Z$ & $A$ & $W$ & $H$ & Total \\[0.1 ex]
		\hline
		u & $1.29\times10^{-12}$ & $29.79 \times 10^{-12} $& $-4.41 \times 10^{-12}$ 
		& $0$ & $26.67 \times 10^{-12}$\\
		d &$5.50 \times 10^{-12}$ & $30.18 \times 10^{-12}$ &
		$ -4.41 \times 10^{-12}$ & $ 0$ & $31.27 \times 10^{-12}$\\
		c & $3.98 \times 10^{-7}$ & $36.91 \times 10^{-7}$ & $-0.48 \times 10^{-7}$ 
		& $0$ & $40.41 \times 10^{-7}$\\ 
		s & $0.22 \times 10^{-8} $& $0.82 \times 10^{-8}$ & $-4.82\times 10^{-8}$ 
		& $ 0 $& $-3.78 \times 10^{-8}$\\
		t & $25.66 \times 10^{-4}$& $10.57 \times 10^{-4}$ & $-2.87 \times 10^{-4}$ 
		& $149.91 \times 10^{-4}$ & $183.27 \times 10^{-4}$\\
		b & $4.16 \times 10^{-6}$ & $7.14 \times 10^{-6}$ & $-98.72 \times 10^{-6}$ 
		& $0.03 \times 10^{-6}$ & $-87.39 \times 10^{-6}$ \\ [1 ex]
		\hline
	\end{tabular}}
		\label{table:weak-MZ}
\end{table} 
\begin{table} [htbp]
	\centering
	%\begin{tabular}{c p{12 mm} c p{15 mm} c p{16 mm} c p{12 mm} c p{12
	%mm} c p{12 mm}} 
	\caption{Electroweak contribution to $4m F_{2}(q^{2}= - m_t^{2})$ 
		of each quark}
	{	\begin{tabular}{cccccc}
		\hline\hline
		Quark & $Z$ & $A$ & $W$ & $H$ & Total \\[0.1 ex]
		\hline
		u & $0.98\times10^{-12}$ & $8.87 \times 10^{-12} $& $-3.30 \times 10^{-12}$ 
		& $0$ & $26.55 \times 10^{-12}$\\
		d &$4.05 \times 10^{-12}$ & $9.03 \times 10^{-12}$ &
		$ -3.30 \times 10^{-12}$ & $ 0$ & $9.78 \times 10^{-12}$\\
		c & $3.01 \times 10^{-7}$ & $11.92 \times 10^{-7}$ & $-0.36 \times 10^{-7}$ 
		& $0$ & $14.57 \times 10^{-7}$\\ 
		s & $1.58 \times 10^{-9} $& $2.53 \times 10^{-9}$ & $-36.25\times 10^{-9}$ 
		& $ 0 $& $-32.14 \times 10^{-9}$\\
		t & $24.54 \times 10^{-4}$& $9.64 \times 10^{-4}$ & $-2.88 \times 10^{-4}$ 
		& $140.72 \times 10^{-4}$ & $172.02 \times 10^{-4}$\\
		b & $3.06 \times 10^{-6}$ & $2.43 \times 10^{-6}$ & $-99.23 \times 10^{-6}$ 
		& $0.02 \times 10^{-6}$ & $-93.72 \times 10^{-6}$ \\ [1 ex]
		\hline
	\end{tabular}}
	\label{table:weak-mt}
\end{table} 
The weak contributions to $4mF_2$ at scale $M_Z$ and at scale $m_t$
 are shown for each quark in Table~\ref{table:weak-MZ} and Table~\ref{table:weak-mt} 
 respectively. In both of these tables, any number which is smaller by at least a 
factor of $10^{-3}$ than the largest number in the same row has been set to zero. 
The factor of $4m$ is conventional, as the quantity $4m F_2$ is dimensionless. 
The total value of $4mF_2 (q^2=- M_{Z}^2)$  for each quark is plotted against 
gluon mass in Fig.~\ref{fig.total_topo_mz}. For comparison, 
we have also plotted the same quantity for each quark when the gluon mass is taken 
to come from a Proca term, as was calculated in~\cite{Choudhury:2014lna}. 
These are shown as dotted lines in these plots.
 Finally, in 
 Fig.~\ref{fig.total_topo_mz_mt} we
 have plotted $4mF_2$ for each quark, at both the scales $m_{t}$ and $M_{Z}$,  
 against topologically generated gluon mass.  
%
%%%%%%%%%%%%%%%%%%%%%%%%%%%%%%%%%%%%%%%%%%
\section{Results and Discussions}
%%%%%%%%%%%%%%%%%%%%%%%%%%%%%%%%%%%%%%%%%%
In this paper we have considered a specific model of gauge-invariant 
mass generation, namely the topologically
massive gauge theory, and calculated the anomalous chromomagnetic moment $F_2(q^2)\,,$ 
at the energy scale $M_Z$ as well as at $m_t$\,. 
Looking at Fig.~\ref{fig.total_topo_mz} we see that gluon mass dependence of the ACM is 
the most prominent for the strange quark. As the gluon mass is increased from 0 to 10 MeV, 
the dimensionless quantity $4m F_2(q^2 = -M_Z^2)$ varies by more than $100\%$ for the $s$-quark 
when the gluon mass is topologically generated. For a Proca mass term this variation 
is only about $42\%$. On the other hand, for the top and bottom quarks 
the mass dependence of the ACM is negligible over this mass range,
irrespective of the mechanism responsible for mass generation (for the $b$ quark the variation is about 0.1\%, for the $t$ quark even less). Among the heavy quarks, 
the charm quark shows approximately $0.5\%$ variation, almost the same for both topological 
and Proca mass terms. For the up and down quarks, the ACM varies by about 20 -- 25\% when the gluon mass is varied over 0 -- 10 MeV, 
and is higher for the topological mass term by about 2\% at the top of the range.

Next we take a look at Fig.~\ref{fig.total_topo_mz_mt}. Here we have plotted the dimensionless 
quantity $4mF_2$ of quarks at two different values of $q^2$ for the same range of topologically 
generated gluon mass. We have plotted $4mF_2$ at the scale $m_t$ as continuous lines, and  
$4mF_2$ at the scale $M_Z$ as dotted lines, against gluon mass between 0 and 10 MeV. From these 
plots, we see that the gluon mass dependence of $4mF_2$ is more pronounced at $q^2=-M^2_Z$ 
than at $q^2=-m^2_t$ for the light quarks like up, down and strange.
 For the top and bottom quarks, the gluon 
mass dependence is negligible for both the energy scales $q^2=-M_Z^2$ and $q^2 = -m_t^2$ 
although the actual values of the ACM are largely different at the two energy scales.

What we can conclude from all this is that the ACM of the light quarks have significant, and 
possibly observable, dependence on gluon mass, irrespective of how the mass is generated.
As higher energies and luminosities become accessible to the LHC, precision measurements 
of the anomalous chromomagnetic moments of quarks should become possible. While a measurement 
of the top quark ACM is unlikely to constrain the gluon mass, data for other quarks will 
be able to put bounds on the mass of the gluon. 

%\newpage

\appendix
%%%%%%%%%%%%%%%%%%%%%%%%%%%%%%%%%%
\section{Calculations}
%%%%%%%%%%%%%%%%%%%%%%%%%%%%%%%%%
This section contains some details of the calculations for Eq.~(\ref{Gamma1})-(\ref{Gamma6}). 
Let us start with $\Gamma^{(1)}_{\mu}$\, as given in Eq.~(\ref{Gamma1}). 
We can write it in the form 
\begin{align}
	&\Gamma^{(1)}_{\mu} = 3!  \int \limits_{0}^{1} d \zeta_{1} \int \limits_{0}^{1} d \zeta_{2}\int \limits_{0}^{1} d \zeta_{3} \int \limits_{0}^{1} d \zeta_{4} \int \dfrac{d^{4}k}{(2 \pi)^4} \delta (1-\zeta_{1} - \zeta_{2} -\zeta_{3} - \zeta_{4})\nonumber\\
	&  \times \dfrac{N^{(1)}_{\mu} (k+ \zeta_{2} p^{\prime}+ (\zeta_{3}+ \zeta_{4})p)}{[k^2 - (\zeta_{2}+\zeta_{3} +\zeta_{4})^2 m^2 + \zeta_{2}(\zeta_{3}+\zeta_{4})q^2 + (-\zeta_{1}+\zeta_{2}+\zeta_{3}+\zeta_{4})m^2 - (\zeta_{2}+\zeta_{3}) M^2]^4}\,,
\end{align}
where we have defined
\begin{align}
	N_{\mu}^{(1)}(k) = (p-k)_{\mu} \gamma_{\lambda} (\slashchar{k} +m) \gamma ^{\lambda} -  (\slashchar{p}-\slashchar{k})(\slashchar{k} +m)\gamma_{\mu}\,.
\end{align}
Similarly, we can write $\Gamma^{(2)}_{\mu}$ as
\begin{align}
	&\Gamma^{(2)}_{\mu} = 3! \int \limits_{0}^{1} d \zeta_{1} \int \limits_{0}^{1} d \zeta_{2}\int \limits_{0}^{1} d \zeta_{3} \int \limits_{0}^{1} d \zeta_{4} \int \dfrac{d^{4}k}{(2 \pi)^4} \delta (1-\zeta_{1} - \zeta_{2} -\zeta_{3} - \zeta_{4})\nonumber\\
	&  \times \dfrac{N^{(2)}_{\mu} (k+ \zeta_{2}p+ (\zeta_{3} + \zeta_{4})p^{\prime})}{[k^2 - (\zeta_{2}+\zeta_{3} +\zeta_{4})^2 m^2 + \zeta_{2}(\zeta_{3}+\zeta_{4})q^2 + (-\zeta_{1}+\zeta_{2}+\zeta_{3}+\zeta_{4})m^2 - (\zeta_{2}+\zeta_{3}) M^2]^4}\,,
\end{align}
with
\begin{align}
	N_{\mu}^{(1)}(k) = \gamma_{\mu} (\slashchar{k} +m)(\slashchar{k}- \slash{p^{\prime}}) - \gamma_{\lambda} (\slashchar{k}+m)\gamma^{\lambda} (k- p^{\prime})_{\mu}\,.
\end{align}
Not all terms of this expression will contribute to $F_2$. Keeping only the relevant terms of
$ N_{\mu}^{(1)}(k)$ and $ N_{\mu}^{(2)}(k)$\,, 
we find that the sum of the contributions from $\Gamma^{(1)}_{\mu}$ and $\Gamma^{(2)}_{\mu}$ is
\begin{align}
	F^{(1)+(2)}_{2} (q^2) =
	%%%2 \times 3! \int \limits_{0}^{1} d \zeta_{1} \int \limits_{0}^{1} d \zeta_{2}\int \limits_{0}^{1} d \zeta_{3} \int \limits_{0}^{1} d \zeta_{4} \int \dfrac{d^{4}k}{(2 \pi)^4} \delta (1-\zeta_{1} - \zeta_{2} -\zeta_{3} - \zeta_{4})\nonumber\\
	%%%&  \times \dfrac{(1- \zeta_{2}- \zeta_{3} - \zeta_{4} )^2}{[k^2 - (\zeta_{2}+\zeta_{3} +\zeta_{4})^2 m^2 + \zeta_{2}(\zeta_{3}+\zeta_{4})q^2 + (-\zeta_{1}+\zeta_{2}+\zeta_{3}+\zeta_{4})m^2 - (\zeta_{2}+\zeta_{3}) M^2]^4} \nonumber\\
	\frac{i}{8\pi ^2 m^3}  \int \limits_{0}^{1} d \zeta_{1} \int \limits_{0}^{1-\zeta_{1}} 
	d \zeta_{2}\int \limits_{0}^{1-\zeta_{1}-\zeta_{2}} d \zeta_{3}\dfrac{\zeta_{1}^{2}}
	{\left[\zeta_{1}^2 - \zeta_{2}(1-\zeta_{1}+\zeta_{2})\frac{q^2}{m^2} - (\zeta_{2}+\zeta_{3})^2
		\frac{M^2}{m^2}\right]^2}\,.
	\label{F2q^2.topoa+b} 
\end{align}

Next we consider $\Gamma^{(3)}_{\mu}$. We can write it as
\begin{align}
	\Gamma^{(3)}_{\mu} &= \frac{2 }{q^2}  \int \limits_{0}^{1} d \zeta_{1} \int \limits_{0}^{1} d \zeta_{2}\int \limits_{0}^{1} d \zeta_{3}  \int \dfrac{d^{4}k}{(2 \pi)^4} \delta (1-\zeta_{1} - \zeta_{2} -\zeta_{3})\nonumber\\
	& \qquad \times \dfrac{N_{\mu}^{(3)}(k+\zeta_{2} p+\zeta_{3} p^{\prime})}{[k^2 - (\zeta_{2}
		+\zeta_{3})^2 m^2 +\zeta_{2}\zeta_{3} q^2 +(-\zeta_{1}+\zeta_{2} + \zeta_{3})m^2 -(\zeta_{2}+\zeta_{3})M^2]^3}\,,
\end{align}
where we have written
\begin{align}
	N_{\mu}^{(3)}(k)= \gamma_{\mu}(\slashchar{k}+m)\slashchar{q}- \slashchar{q}(\slashchar{k}+m)\gamma_{\mu}.
	\label{Numerator3}
\end{align}
The contribution for $ \Gamma^{(3)}_{\mu} $ is easily calculated from this to be
\begin{align}
	F^{(3)}_{2}(q^2) 
	%%% &= -\frac{4m}{q^2}\int \limits_{0}^{1} d \zeta_{1} \int \limits_{0}^{1} d \zeta_{2}\int \limits_{0}^{1} d \zeta_{3}  \int \dfrac{d^{4}k}{(2 \pi)^4} \delta (1-\zeta_{1} - \zeta_{2} -\zeta_{3})\nonumber\\ 
	%%% & \qquad \times \dfrac{(1-\zeta_{2}-\zeta_{3})}{[k^2 - (\zeta_{2}+\zeta_{3})^2 m^2 +\zeta_{2}\zeta_{3} q^2 +(-\zeta_{1}+\zeta_{2}+ \zeta_{3})m^2 -(\zeta_{2}+\zeta_{3})M^2]^3}\nonumber\\
	= \frac{i}{8 \pi^2 mq^2} \int\limits_{0}^{1} d \zeta_{1} \int \limits_{0}^{1-\zeta_{1}} 
	d\zeta_{2}\dfrac{\zeta_{1}}{\zeta_{1}^2 -(1-\zeta_{1}-\zeta_{2})\zeta_{2}\frac{q^2}{m^2}
		+ (1-\zeta_{1} )\frac{M^2}{m^2}}\,.
	\label{F2Q^2.topoc}
\end{align}
The remaining three $\Gamma$'s have long expressions. We will show the calculation of $\Gamma^{(4)}_\mu$ in some detail, showing only the final expression for the other two. We can write $\Gamma^{(4)}_\mu$ as
\begin{align}
	\Gamma^{(4)}_{\mu} = 4! \int \limits_{0}^{1} d \zeta_{1} \int \limits_{0}^{1} d \zeta_{2}\int \limits_{0}^{1} d \zeta_{3} \int \limits_{0}^{1} d \zeta_{4} \int \limits_{0}^{1} d \zeta_{5} \int \dfrac{d^{5}k}{(2 \pi)^4} \delta (1-\zeta_{1} - \zeta_{2} -\zeta_{3} - \zeta_{4}-\zeta_{5})\nonumber\\
	\times \dfrac{N_{\mu}^{(4)}\left(k+(\zeta_{2}+\zeta_{3})p +(\zeta_{4}+\zeta_{5})p^{\prime}\right)}{[k^2 - (\zeta_{2}+\zeta_{3}+\zeta_{4}+\zeta_{5})^2 m^2 + (\zeta_{2}+\zeta_{3})(\zeta_{4}+\zeta_{5})q^2 +(\zeta_{2}+\zeta_{3 } +\zeta_{4} +\zeta_{5}-\zeta_{1})m^2- (\zeta_{2}+\zeta_{4})M^2]^5}\,,
\end{align}
where the function in the numerator is 
\begin{align}
	N_{\mu}^{(4)}(k) = &-(\slashchar{k}- \slashchar{p})(\slashchar{k}+ m)(\slashchar{k}-\slashchar{p^{\prime}})(2k-p-p^{\prime})_{\mu} +2(k-p)_{\mu} (\slashchar{k} - \slashchar{p^{\prime}})(\slashchar{k}+m)(\slashchar{k}-\slashchar{p^{\prime}})\nonumber\\
	& + 2(k-p^{\prime})_{\mu} (\slashchar{k} - \slashchar{p})(\slashchar{k}+m)(\slashchar{k}-\slashchar{p})
	- (k-p^{\prime}) \cdot (k-p) \lbrace 2(\slashchar{k}- \slashchar{p^{\prime}})(\slashchar{k}+m) \gamma_{\mu}\nonumber\\
	& - \gamma_{\lambda} (\slashchar{k}+m) \gamma^{\lambda} (2k-p-p^{\prime})_{\mu} + 2\gamma_\mu (\slashchar{k} +m) (\slashchar{k} - \slashchar{p})\rbrace\,.
	\label{numerator4}
\end{align}
Changing variables from $k$ to $k+(\zeta_{2}+\zeta_{3})p +(\zeta_{4}+\zeta_{5})p^{\prime}\,,$ we can write the first term in Eq.~(\ref{numerator4}) as
\begin{equation}
- (\slashchar{k} + \slashchar{a} -\slashchar{p})(\slashchar{k}+\slashchar{a}+ m)(\slashchar{k}+\slashchar{a}-\slashchar{p^{\prime}})\lbrace 2k + (2\zeta_{2}+2\zeta_{3}-1)p+ (2\zeta_{3}+2 \zeta_{4}-1)p^{\prime}\rbrace _{\mu}\,,
\label{numerator4.term1}
\end{equation}
where we have defined
\begin{equation}
	a_{\mu} = (\zeta_{2}+ \zeta_{3}) p_{\mu} +(\zeta_{3}+\zeta_{4})p^{\prime} _{\mu}\,.
\end{equation}
As before, we ignore terms in Eq.~(\ref{numerator4.term1}) which do not contribute
to $F_{2}(q^2)$. The relevant terms can then be written as 
\begin{align}
	&-\frac{1}{2}mk^2\left({3}(A+B)-2\right) b_{\mu}+mk^2(1-A-B)a_{\mu}\nonumber\\
	& \qquad \qquad + (1-A-B)mb_{\mu}\left[\left( (A+B)^2-1 \right) m^{2} + (1-AB)q^2\right]\,,
	\label{numerator4.relevant_term1}
\end{align}
where we have written
\begin{align}
	A= \zeta_{2}+ \zeta_{3}\,,\qquad
	B=\zeta_{4} +\zeta_{5}\,,
\end{align}
and
\begin{align}
	b_{\mu} = (2A-1)p_{\mu}+ (2B-1)p^{\prime}_{\mu}\,.
\end{align}
The second and third terms in Eq.~(\ref{numerator4}), when added together, produce
\begin{align}
	2(\slashchar{k}-m)(k^{2}-m^2)(2k- p-p^{\prime})\,.
	\label{numerator4.term2+3}
\end{align}
After transforming from $k$ to $k+ (\zeta_{2}+\zeta_{3})p +(\zeta_{4}+\zeta_{5})p^{\prime}$\,, we can write the relevant terms in Eq.~(\ref{numerator4.term2+3}) as
\begin{align}
	-mk^{2}a_{\mu}(1-A-B) &+ mb_{\mu} [k^{2}\left((A+B)-2\right)\nonumber\\ &+m^2 (1-A-B)^2(1+A+B)
	+AB(1-A-B)q^2]\,.
	\label{numerator4.term2+3rel}
\end{align}
It is easy to see that the fourth and the last terms in Eq.~(\ref{numerator4})
do not contribute. The relevant expression contributed by the fifth term of 
Eq.~(\ref{numerator4}) are
\begin{align}
	[-2k^2  - 2m^2(1-A-B)^2 + \left( AB+(A-1)(B-1)\right) q^2 ](A+B-2)mb_{\mu}\,.
	\label{numerator4.term5rel}
\end{align}
Adding Eq.~(\ref{numerator4.term1}),~(\ref{numerator4.term2+3rel}) and~(\ref{numerator4.term5rel})and their forms obtained by interchanging dummy variables $\zeta_{2}$, $\zeta_{4}$ and $\zeta_{3}$, $\zeta_{5}$, we get from $N_{\mu}^{(4)}(k)$\,,
\begin{align}
	&\left[-3(1-A-B)mk^2 -2(1-A-B)^3(2-A-B)m^3\right.\nonumber\\
	& \qquad \left. +\left((1-A-B)^3 +2AB(1-A-B)(2-A-B)\right) mq^2\right](p+p^{\prime})_{\mu}\,.
\end{align}
Using Gordon's identity, we find that  the contribution to $F_2(q^2)$ from $\Gamma^{(4)}_{\mu}$ can be written as
\begin{align}
	F^{(4)}_{2}(q^2) =& -\dfrac{i}{8 \pi^2 m^3} \int \limits_{0}^{1} d \zeta_{1} \int \limits_{0}^{1}d\zeta_{2} \int \limits_{0}^{1} d \zeta_{3} \int \limits_{0}^{1} d \zeta_{4} \int \limits_{0}^{1}d\zeta_{5}\,  \delta(1-\zeta_{1}-\zeta_{2}-\zeta_{3}-\zeta_{4}-\zeta_{5})\nonumber\\ 
	&  \times\left[ \dfrac{3(1-\zeta_{2}-\zeta_{3}-\zeta_{4}-\zeta_{5})}{\left[ (1-\zeta_{2}-\zeta_{3}-\zeta_{4}-\zeta_{5})^2  -(\zeta_{2} +\zeta_{3})(\zeta_{4}+\zeta_{5})\frac{q^2}{m^2} +(\zeta_{2}+\zeta_{4})
		\frac{M^2}{m^2}\right]^2}\right. \nonumber\\
	& +  \dfrac{2(1-\zeta_{2}-\zeta_{3}-\zeta_{4}-\zeta_{5})^3 (\zeta_{2}+\zeta_{3}+\zeta_{4}+\zeta_{5}-2)}{\left[ (\zeta_{2}+\zeta_{3}+\zeta_{4}+\zeta_{5})^2  -(\zeta_{2} +\zeta_{3})(\zeta_{4}+\zeta_{5})\frac{q^2}{m^2} -(\zeta_{2} +\zeta_{3} +\zeta_{4} +\zeta_{5} -\zeta_{1})+(\zeta_{2}+\zeta_{4})\frac{M^2}{m^2}\right]^3} \nonumber\\
	&\left.+\dfrac{q^2}{ m^2} \dfrac{(1-\zeta_{2}-\zeta_{3}-\zeta_{4}-\zeta_{5})\left[ 2(2-\zeta_{2}-\zeta_{3}-\zeta_{4}-\zeta_{5}) (\zeta_{2}+\zeta_{3})(\zeta_{4}+\zeta_{5}) + (1-\zeta_{2}-\zeta_{3}-\zeta_{4}-\zeta_{5})^2\right]}{\left[ (\zeta_{2}+\zeta_{3}+\zeta_{4}+\zeta_{5})^2  -(\zeta_{2} +\zeta_{3})(\zeta_{4}+\zeta_{5})\frac{q^2}{m^2} -(\zeta_{2} +\zeta_{3} +\zeta_{4} +\zeta_{5} -\zeta_{1})+(\zeta_{2}+\zeta_{4})\frac{M^2}{m^2}\right]^3}\right]\,.
	\label{F2q^2.topod}
\end{align}
%
%\warning{last numerator rewritten. check!}

We can calculate the contributions from $\Gamma^{(5)}_{\mu}$ and $\Gamma^{(6)}_{\mu}$ in a similar manner, and their sum has a fairly simple expression, 
\begin{align}
	F^{(5)+(6)}_{2}(q^2) = &-\dfrac{i}{4 \pi ^2 m^3} \int \limits_{0}^{1} d \zeta_{1} \int \limits_{0}^{1} d \zeta_{2}\int \limits_{0}^{1} d \zeta_{3} \int \limits_{0}^{1} d \zeta_{4} \, \delta(1-\zeta_{1}-\zeta_{2}-\zeta_{3}-\zeta_{4})\nonumber\\
	& \times \dfrac{\zeta_{2}(\zeta_{2}+\zeta_{3}+\zeta_{4}-1)}{\left[(\zeta_{2}+\zeta_{3}+\zeta_{4})^2 - \zeta_{2}(\zeta_{3}+\zeta_{4}) \frac{q^2}{m^2} -(-\zeta_{1}+\zeta_{2}+\zeta_{3}+\zeta_{4}) +(\zeta_{2}+\zeta_{3})\frac{M^2}{m^2}\right]^2}\,.
	\label{F2q^2.topoe+f}
\end{align}
%

%%%%%%%%%%%%%%%%%%%%%%%%%%%%%%%%%%%%%%%%%%%%% 
{}

\end{document}